\documentclass[aps, prd, english, superscriptaddress, longbibliography, 11pt, notitlepage, nofootinbib]{revtex4}

\immediate\write16{<<WARNING: LINEDRAW macros work with emTeX-dvivers and other drivers supporting emTeX \special's (dviscr, dvihplj, dvidot, dvips, dviwin, etc.)>>}

\usepackage[colorlinks=true, a4paper=true, pdfstartview=FitV, linkcolor=blue, citecolor=blue, urlcolor=blue]{hyperref}
\usepackage{caption}
\usepackage{subcaption} 
\usepackage{amsmath}
\usepackage{physics}
\usepackage{amssymb}
\usepackage{graphicx}
\usepackage{hhline}
\usepackage{mwe}
\usepackage{amsbsy}
\usepackage{textcomp}
\usepackage{commath}

\usepackage{slashed}
\usepackage{natbib}

\usepackage{stackengine}
\stackMath

\makeatletter

\usepackage{babel}
\numberwithin{equation}{section}

\begin{document}

\title{Vortex-Antivortex Annihilation from BPS Equations}
% Multi-vortex systems as solutions of a mirror-impurity modified Abelian-Higgs model
%\preprint{KA-TP-??-2017}
%\date{23-03-2022}

\author{J. P. S. Neto}
\email{jose.neto@ufpe.br}
\affiliation{Departamento de Física, Universidade Federal de Pernambuco, Av. Prof. Moraes Rego, 1235, Recife - PE - 50670-901, Brazil}

%\author{F. R. Silva}
%\email{filipe.rodrigues@ufpe.br}
%\affiliation{Departamento de Física, Universidade Federal de Pernambuco, Av. Prof. Moraes Rego, 1235, Recife - PE - 50670-901, Brazil}

\author{J. G. F. Campos}
\email{joao.gfcampos@upe.br}
\affiliation{Física de Materiais, Universidade de Pernambuco, Rua Benfica, 455, Recife - PE - 50720-001, Brazil}

\author{A. Mohammadi}
\email{azadeh.mohammadi@ufpe.br}
\affiliation{Departamento de Física, Universidade Federal de Pernambuco, Av. Prof. Moraes Rego, 1235, Recife - PE - 50670-901, Brazil}

\begin{abstract}

In this work, we introduce a vortex-impurity model that admits BPS
configurations containing both vortices and antivortices and allows
vortex-antivortex annihilation to be described within the moduli-space
approximation. For a suitable class of impurities, the BPS equations
can be mapped through a coordinate transformation onto those of the
impurity-free Abelian-Higgs model. In particular, we identify
nonlocalized impurities that act as mirrors, generating a vortex-antivortex pair symmetrically placed with respect to the impurity. We derive the corresponding moduli-space metric and show that, within the geodesic approximation, the pair approaches the annihilation configuration asymptotically. We further extend the construction to chains and two-dimensional arrays of alternating vortices and antivortices. Finally, we introduce radial bump impurities that produce stretched and compressed vortices with nontrivial internal magnetic structures.
   
\end{abstract}

\maketitle

%%%%%%%%%%%%%%%%%%%%%%%%%%%%%%%%%%%%%%%%%%%%%%%%%
\section{Introduction}\label{intro} \label{sec:intro}

Heuristically, two ingredients are required for a field theory to support vortex solutions \cite{manton}. First, there must exist an isolated point $p$ where the scalar field $\phi$ vanishes. Second, the phase of $\phi$ must wind by an integer multiple of $2\pi$ along a closed curve around $p$. Such configurations were first predicted by Abrikosov in his pioneering work on the transition between type-I and type-II superconductors \cite{abrikosov}, within the framework of the Ginzburg-Landau theory \cite{ginzburg}. In 1973, Nielsen and Olesen showed that relativistic vortex solutions also arise in the Abelian Maxwell-Higgs (AMH) model \cite{nielsen}, a relativistic extension of the Ginzburg-Landau theory in which the Higgs scalar field $\phi$ is coupled to gauge fields $A_{\mu}\;(\mu=0,1,2)$. Since then, vortex solutions have been studied in many other theories. Examples include the Abelian Chern-Simons (ACS) model \cite{acs1, acs2, acs3}, which also supports gauged vortices, and the gravitating AMH model, whose vortex solutions correspond to cosmic strings \cite{vilenkin}.

Additionally, vortices belong to the broader class of topologically stable configurations known as \textit{solitons}, which arise in field theories with spontaneous symmetry breaking. The stability of these particle-like solutions was investigated by Bogomol'nyi \cite{bogomolnyi}, who showed that the energy is minimized when the fields satisfy a set of first-order differential equations. These are the so-called \textit{Bogomol'nyi equations}, or \textit{Bogomol'nyi-Prasad-Sommerfield (BPS) equations} \cite{prasad}. Static kinks, critically coupled vortices, and BPS monopoles are well-known examples of solutions satisfying the BPS equations of their respective field theories.

The space of all solutions of the BPS equations,  $\mathcal{M}_N$, labeled by a set of $N$ moduli parameters, is called the \textit{moduli space}. Since all BPS solutions saturate the same energy bound, there is no static force between them. Nevertheless, the moduli space can have a nontrivial geometry, giving rise to an effective interaction through the curvature of the manifold $\mathcal{M}_N$ \cite{manton1982}.

Many studies have explored the effects of impurities on the moduli space of BPS solitons. For example, ref.~\cite{adam2019} introduced a $(1+1)$-dimensional kink model in which the impurity induces a nontrivial moduli-space metric that can be computed analytically. As a result, the low-velocity dynamics is described by geodesic motion, since the interaction potential vanishes for BPS configurations. This framework was then applied to kink-antikink collisions. In the absence of impurities, a kink-antikink pair cannot form a BPS solution because of their attractive interaction. However, with a suitable choice of impurity, both kink and antikink become BPS solutions and annihilate through geodesic motion on the moduli space.

A natural question is whether analogous results can be obtained in a vortex-impurity model. Several higher-dimensional theories admit BPS soliton-impurity interactions. For magnetic impurities, this has been demonstrated in the Abelian-Higgs model \cite{tong, cockburn2017, han2016} and for Chern-Simons vortices \cite{han2016, bazeia2024}. Other works have used auxiliary functions to construct BPS vortices \cite{bazeia2018, bazeia2018internal, bazeia2019multi} and BPS monopoles \cite{bazeia2018monopoles} with nontrivial internal structure. Coexisting vortex-antivortex (v-av) BPS solutions have also been reported in a product Abelian-Higgs theory with impurities \cite{han2021}.

In general, the dynamics of v-av is an important topic in soliton physics. For example, ref.~\cite{gleiser2007} showed that low-momentum vortex-antivortex collisions can produce oscillons as long-lived remnants. More recently, ref.~\cite{bachmaier2026} reported rich chaotic dynamics in collisions between excited vortex-antivortex pairs. A moduli-space description of these systems may provide valuable insight into their dynamics while greatly simplifying the analysis.

In this paper, we present a vortex-impurity model that admits BPS equations and an associated moduli space by construction. Following the approach of refs.~\cite{adam2019, manton2019, slawinska2025}, we construct vortex-antivortex pairs that annihilate through moduli-space dynamics. We also introduce an alternative method for generating vortices with different internal structures using generalized bump impurities, first proposed in ref.~\cite{manton2019}.

The paper is organized as follows. In section~\ref{sec:2}, we briefly
review the AMH model and introduce the BPS-impurity vortex equations.
In section~\ref{sec:3}, we develop a method for solving these equations,
construct vortex-antivortex configurations, and introduce an iterative scheme for generating more general arrangements of vortices and
antivortices. In section~\ref{sec:4}, we construct radially symmetric solutions in the presence of bump impurities and study their internal magnetic structure. In section~\ref{sec:5}, we derive the Lagrangian
formulation of the BPS-impurity model. In section~\ref{sec:6}, we study the low-velocity dynamics of the vortex-antivortex pair using the
geodesic approximation and discuss the difficulties associated with
extending this analysis to the full second-order dynamics. Finally, in
section~\ref{sec:conc}, we summarize our main results and discuss possible directions for future work.

%%%%%%%%%%%%%%%%%%%%%%%%%%%%%%%%%%%%%%%%%%%%%%%%%%%%%%%%%%%%%%%%%%%
\section{BPS-impurity equations} \label{sec:2}

The Nielsen-Olesen vortex arises as a static solution of the AMH model, a $(2+1)$-dimensional field theory described by the Lagrangian density
\begin{align}
    \mathcal{L} = -\frac{1}{4}F^{\mu \nu}F_{\mu \nu} 
    + \frac{1}{2}D_{\mu} \phi \overline{D^{\mu} \phi}  
    - \frac{\lambda}{8}\left(|\phi|^2 - 1\right)^2,
    \label{eq:2.1}
\end{align} 
where $D_\mu = \partial_\mu - iA_\mu$ is the covariant derivative and $F_{\mu \nu} = \partial_\mu A_\nu - \partial_\nu A_\mu$ is the field strength tensor $(\mu,\nu=0,1,2)$. In the static limit, the corresponding energy functional reduces to the Ginzburg-Landau energy,
\begin{align}
    E = \frac{1}{2}\int d^2x \left\{
    B^2 + D_i \phi \overline{D_i \phi} 
    + \frac{\lambda}{4}\left(|\phi|^2 - 1\right)^2 
    \right\} , \label{eq:2.2}
\end{align}
where $B=\partial_1 A_2 - \partial_2 A_1$ is the only nonvanishing component of the magnetic field. This model has $U(1)$ gauge symmetry, which naturally leads to the use of polar coordinates $(r,\theta)$ and a radially symmetric ansatz for the vortex solutions. For the vortex to be a localized finite-energy configuration, the fields must satisfy
\begin{equation}
|\phi|\to1,\qquad
B\to0,\qquad
D_i\phi\to0,
\qquad \text{as }\, r\to\infty,
\label{eq:2.3}
\end{equation}
so that the Higgs field approaches its vacuum expectation value and reduces to the pure phase
$\phi(\infty)=e^{i\varphi}$, where $\varphi$ is the phase in the internal space. Consequently, $\phi(\infty)$ defines a map $S^1_{\infty}\rightarrow S^1$, relating the circle at spatial infinity to the vacuum manifold $|\phi|^2=1$ in the internal $U(1)$ space.
These maps are classified by the winding number $N$, an integer-valued topological invariant defined by
\begin{align}
    N=\frac{1}{2\pi}\int_0^{2\pi}d\theta\,\partial_\theta\varphi =\frac{1} {2\pi}\left[\varphi(2\pi)-\varphi(0)\right].
    \label{eq:2.4}
\end{align}
For $N\neq0$, continuity of the Higgs field requires $\phi$ to vanish at least once in the plane. Such a zero is known as the vortex center.

Static vortex solutions are obtained by solving the Euler-Lagrange equations derived from eq.~(\ref{eq:2.1}) together with the boundary conditions (\ref{eq:2.3}). At the critical coupling ($\lambda=1$), however, it is sufficient to solve the first-order BPS equations,
\begin{align}
&D_1 \phi \pm i D_2 \phi = 0,\label{eq:2.5}\\
&B \pm \frac{1}{2} \left( |\phi|^2 - 1 \right) = 0 , \label{eq:2.6}
\end{align}
whose solutions minimize the energy functional (\ref{eq:2.2}) and therefore also satisfy the full second-order field equations. They saturate the Bogomol'nyi bound, $E =\frac{1}{2}\int d^2x \; B =  \pi |N|$.

In eqs.~(\ref{eq:2.5})-(\ref{eq:2.6}), the upper sign corresponds to $N>0$ (vortices), while the lower sign corresponds to $N<0$ (antivortices). Taubes showed that these equations admit $N$-vortex solutions whose Higgs-field zeros lie in $\mathbb{R}^2$, each carrying a positive multiplicity such that the sum of all multiplicities equals $N$ \cite{taubes1, taubes2, taubes3}. These solutions form a $2N$-dimensional moduli space, $\mathcal{M}_N\simeq\mathbb{C}^N$, with the vortex positions serving as moduli \cite{samols1992}. By contrast, the critical theory admits no BPS vortex-antivortex solutions, and therefore no moduli space describing their dynamics.

Motivated by these observations, we consider an extension of the Abelian-Higgs model described by the BPS equations
\begin{align}
 &\sigma_1(\mathbf{x})D_1 \phi + i \sigma_2(\mathbf{x}) D_2 \phi = 0, \label{eq:2.7}\\
&\sigma_1(\mathbf{x})\sigma_2(\mathbf{x})B + \frac{1}{2} \left( |\phi|^2 - 1 \right) = 0 ,\label{eq:2.8}
\end{align}
where the impurity functions $\sigma_i=\sigma_i(x,y)$ $(i=1,2)$ couple to both the Higgs and gauge fields. Throughout this work, we assume that the impurities preserve one half of the original BPS structure, as is normally the case in soliton-impurity theories (see, for example, refs. \cite{adam2019, bazeia2024}). The corresponding Lagrangian formulation will be presented in section \ref{sec:5}.

The proposed model admits a new class of BPS solutions. Besides recovering the standard vortices when $\sigma_i=1$, it also supports coexisting vortex-antivortex pairs and compressed or stretched vortices with nontrivial internal structure. We first focus on the construction of v-av pairs by choosing $\sigma_2=1$ and requiring $\sigma_1\rightarrow \pm1$ as $x \rightarrow \pm \infty$. As a result, the BPS equations reduce asymptotically to the antivortex equations on the left and to the vortex equations on the right. This construction naturally gives rise to a v-av pair described by a single modulus parameter.

\section{Mirror impurities} 
\label{sec:3}

To solve the BPS-impurity equations, we introduce a coordinate transformation that maps eqs.~(\ref{eq:2.7})-(\ref{eq:2.8}) onto the original BPS equations (\ref{eq:2.5})-(\ref{eq:2.6}). Consider an invertible transformation $x=x(u,v)$ and $y=y(u,v)$ such that
\begin{equation}
\sigma_1D_1\phi \rightarrow D_u\phi,
\qquad
\sigma_2D_2\phi \rightarrow D_v\phi.
\end{equation}
These conditions are satisfied provided
\begin{align}
    &\partial_x u = 1/\sigma_1 ,  \label{eq:3.1}\\
    &\partial_y u = 0 , \label{eq:3.2} \\
    &\partial_x v = 0 , \label{eq:3.3} \\ 
    &\partial_y v = 1/\sigma_2.\label{eq:3.4}
\end{align}
This coordinate transformation also induces a transformation of the gauge field components,
\begin{equation}
A_u=\sigma_1A_1,\qquad
A_v=\sigma_2A_2,
\end{equation}
so that the magnetic field becomes
\begin{align}
    B_w = \partial_u A_v - \partial_v A_u = \sigma_1 \sigma_2 (\partial_1 A_2 - \partial_2 A_1) = \sigma_1 \sigma_2 B . \label{eq:3.5}
\end{align} 
As a result, the BPS-impurity equations reduce to the original BPS equations in the $(u,v)$ plane,
\begin{align}
&D_u \phi + i D_v \phi = 0, \label{eq:3.6} \\
&B_w + \frac{1}{2} \left( |\phi|^2 - 1 \right) = 0. \label{eq:3.7}
\end{align}

Differentiating eqs.~(\ref{eq:3.1}) and (\ref{eq:3.4}) with respect to $y$ and $x$, respectively, and using eqs.~(\ref{eq:3.2}) and (\ref{eq:3.3}) gives $\partial_y \sigma_1 = 0$ and $\partial_x \sigma_2 = 0$\footnote{This follows from assuming that the impurity functions are sufficiently smooth so that $\partial_x\partial_y\sigma_i=\partial_y\partial_x\sigma_i$ $(i=1,2)$ almost everywhere.}.
These conditions are necessary for the coordinate transformation to map the BPS-impurity equations onto the standard BPS equations. Consequently, each impurity function must depend only on its associated coordinate; otherwise, the construction described in this section is no longer valid.

The coordinate transformation must be invertible in order to solve the equations in the $(u,v)$ plane and map the solutions back to the Cartesian coordinates. By the inverse function theorem, this requires the Jacobian determinant to be nonvanishing. In the present case,
\begin{align}
    J = \frac{\partial(x, y)}{\partial(u, v)} = \begin{pmatrix}
        \partial_u x & \partial_v x \\
        \partial_u y & \partial_v y \\
    \end{pmatrix} = \begin{pmatrix}
        \sigma_1(x) & 0 \\
        0 & \sigma_2(y) \\
        \end{pmatrix} \Rightarrow \det J = \sigma_1(x)\sigma_2(y) ,
        \label{eq:3.8}
\end{align}
Therefore, the transformation ceases to be invertible at points where either impurity function vanishes, and the construction presented here is not applicable in such regions.

The BPS equations (\ref{eq:3.6})-(\ref{eq:3.7}) are solved as in the standard theory. Let us introduce polar coordinates in the $(u,v)$ plane, $\rho = \sqrt{u^2 + v^2}$ and $\vartheta = \arctan(v/u)$, and adopt the vortex \textit{ansatz}
\begin{align}
&\phi(\rho, \vartheta) = f(\rho)e^{i\vartheta} \label{eq:3.9}, \\
&A_u(\rho, \vartheta) = -\frac{v g(\rho)}{\rho^2} \label{eq:3.10} , \\
&A_v(\rho, \vartheta) = \frac{u g(\rho)}{\rho^2} \label{eq:3.11}.
\end{align} 
Substituting this \textit{ansatz} into the BPS equations separates the radial and angular variables, leading to the ODEs
\begin{align}
&f' = \frac{f}{\rho}(1-g), \label{eq:3.12} \\
&g' = \frac{\rho}{2}(1-f^2). \label{eq:3.13}
\end{align} 
These equations are solved numerically with the boundary conditions
$f(0)=g(0)=0$ and $f(\infty)=g(\infty)=1$. Transforming back to Cartesian coordinates using eqs.~(\ref{eq:3.1})-(\ref{eq:3.4}), the Higgs and gauge fields become
\begin{align}
&\phi(x, y) = f(\rho(u, v))e^{i\vartheta(u, v)} \label{eq:3.14}, \\
&A_i(x, y) = -\frac{\epsilon_{ij} \xi^j}{\sigma_i(x^i) \rho^2}g(u, v) \;\; ; \quad \xi^i = (\xi^1, \xi^2) \equiv(u(x), v(y)). \label{eq:3.15}
\end{align} 
The magnetic field then transforms according to eq. (\ref{eq:3.5}) giving rise to
\begin{align}
   B = \frac{(\partial_u A_v - \partial_v A_u)}{\sigma_1(x)\sigma_2(y)} = \frac{1}{\sigma_1(x)\sigma_2(y)}\frac{g'(\rho(u, v))}{\rho}. \label{eq:3.16}
\end{align} 
Equations (\ref{eq:3.14})-(\ref{eq:3.16}) completely determine the solutions of the BPS-impurity equations.

As anticipated in section \ref{sec:2}, we use eqs.~(\ref{eq:3.14})-(\ref{eq:3.16}) to construct vortex-antivortex configurations. In analogous kink-impurity models \cite{adam2019, manton2019}, the choice $\sigma(x)=\tanh x$ generates kink-antikink pairs because of its sigmoid profile. In the vortex case, however, this choice leads to additional complications.

Applying the construction described above to the impurities $\sigma_1(x)=\tanh x$ and $\sigma_2(y)=1$ yields the coordinate transformation
\begin{align}
   (x, y) \to \left(u(x), v(y)\right) = \left(\ln (\frac{|\sinh x|}{\sinh a_x}), y-a_y \right) , \;\;\; \text{with} \;\;\; a_x>0 ,\label{eq:3.17}
\end{align} 
where $(a_x,a_y)$ are the BPS moduli arising as integration constants of eqs.~(\ref{eq:3.1})-(\ref{eq:3.4}). The coordinate $u(x;a_x)$ covers only $\mathbb{R}_{-}$ and becomes singular at $x=0$. A similar issue appears in kink-impurity models with $\sigma(x)=\coth x$ \cite{slawinska2025}. 
Fortunately, standard one-dimensional models, such as the $\phi^4$, sine-Gordon, and $\phi^6$ theories, admit solutions that can be written as functions of $e^{u(x)}$. This removes the singularity of $u$ and keeps the fields well defined at $x=0$.
In ref.~\cite{slawinska2025}, an analytic continuation of the BPS modulus $a_x$ is performed, extending the coordinate $u(x;a_x)$, and hence the Higgs field, to the region $x>0$. With an appropriate choice of branch cut, the analytic continuation produces a kink-kink configuration instead of simply reflecting the solution across $x=0$. In the present case, however, the fields (\ref{eq:3.14})-(\ref{eq:3.16}) depend on $u(x;a_x)$ through the polar coordinates $(\rho,\vartheta)$, making it impossible to remove the singularity in the same way. In this context, no ideal analytical continuation exists in the model.

An alternative choice is $\sigma_1(x)=\coth x$, leading to the transformation
\begin{equation}
    u(x; a_x) = \ln \left(\frac{\cosh x}{\cosh a_x}\right).
\end{equation}
The impurity is singular at $x=0$, which means that $u(x; a_x)$ is not invertible. In fact, the full $x$-axis maps onto only a semi-infinite portion of the $u$-axis. The map is also not injective. Nevertheless, the solution is well-defined. In addition, the singularity imposes the boundary conditions
\begin{align}
D_1 \phi \big|_{x=0} = 0, \quad
B\big|_{x=0} = 0. \label{eq:3.19}
\end{align}
They are satisfied by all BPS solutions, as can be verified directly from the analytical expressions.

The coordinate $u(x; a_x)$
appears to be restricted by the condition $\cosh a_x\ge1$. This restriction can be removed by analytically continuing $a_x$ to purely imaginary values. It is then convenient to introduce the modulus $\alpha_x = \cosh a_x$, which extends the allowed range to $\alpha_x\in(0,1]$, covering the entire moduli space. Throughout this work, we use either parametrization, depending on which is more convenient.

\subsection{$\coth$-Mirror impurities} \label{sec:3.A}

It is straightforward to verify that the impurity $\sigma_1(x)=\coth x$ exhibits the asymptotic behavior anticipated in section~\ref{sec:2}. To locate the av-v pair, note that the vortex solution in the $(u,v)$ plane, centered at the origin, is mapped to the points $(\pm a_x,a_y)$ in the Cartesian plane. These points correspond to the antivortex (minus sign) and vortex (plus sign) centers. The symmetry between their positions is expected, since a first-order differential equation has at most one free parameter in each spatial direction. Consequently, the pair is constrained to emerge symmetrically with respect to the impurity, which motivates the name of this section.

\begin{figure}[t!]
    \centering
    % Primeira linha (2 figuras)
    \begin{subfigure}{0.48\textwidth}
        \includegraphics[width=\linewidth]{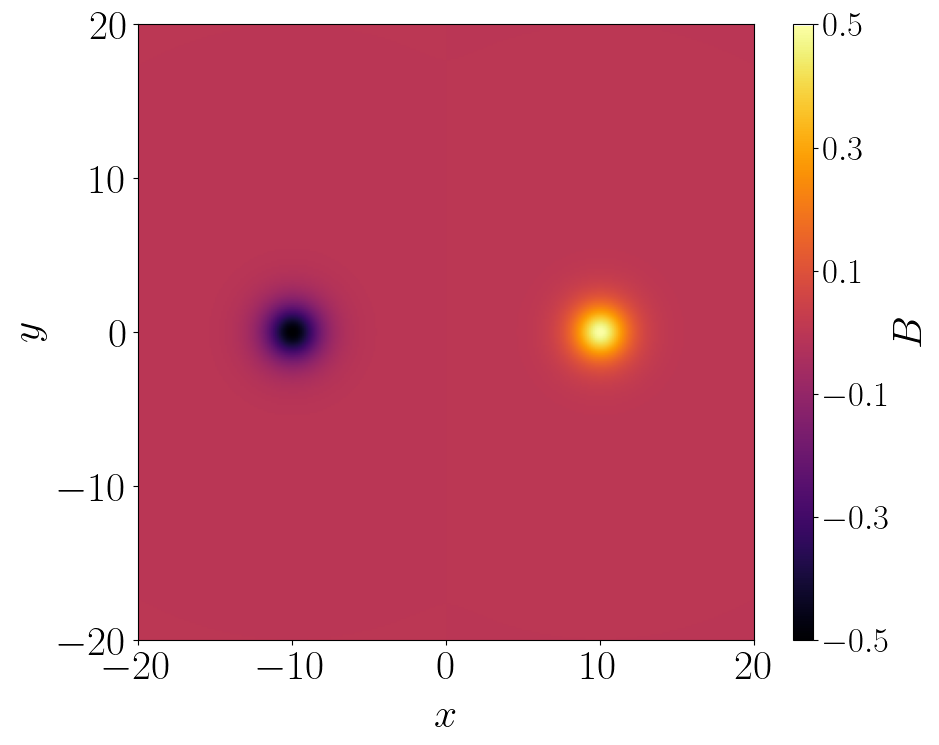}
        \caption{}
    \end{subfigure}
    \hfill
    \begin{subfigure}{0.48\textwidth}
        \includegraphics[width=\linewidth]{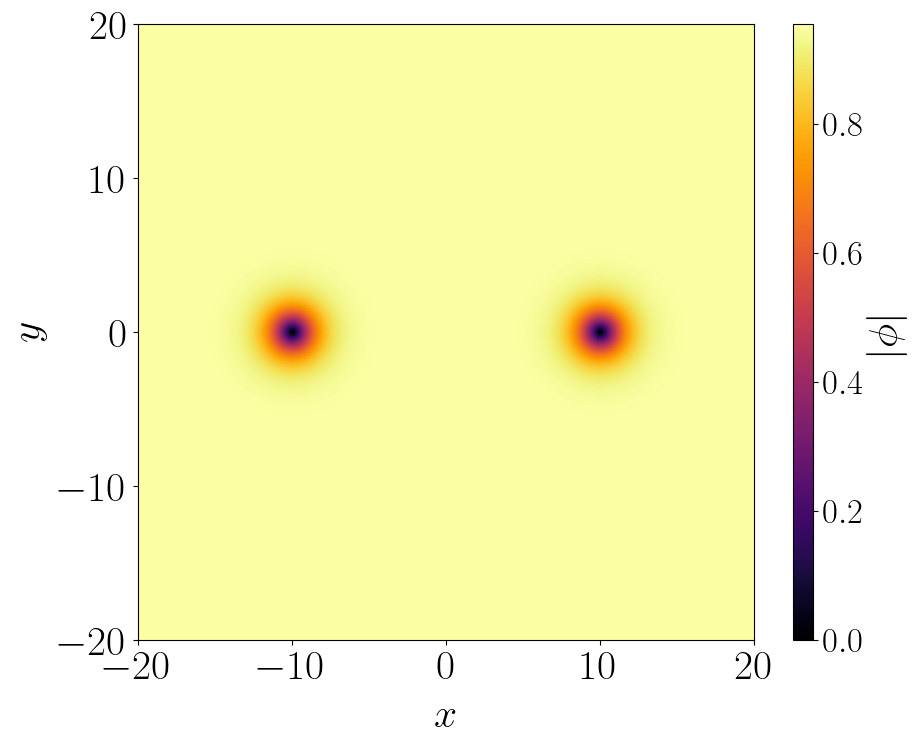}
        \caption{}
    \end{subfigure}
    
    \vspace{0.01cm} % Espaço vertical entre as linhas
    
    % Segunda linha (2 figuras)
    \begin{subfigure}{0.48\textwidth}
        \includegraphics[width=\linewidth]{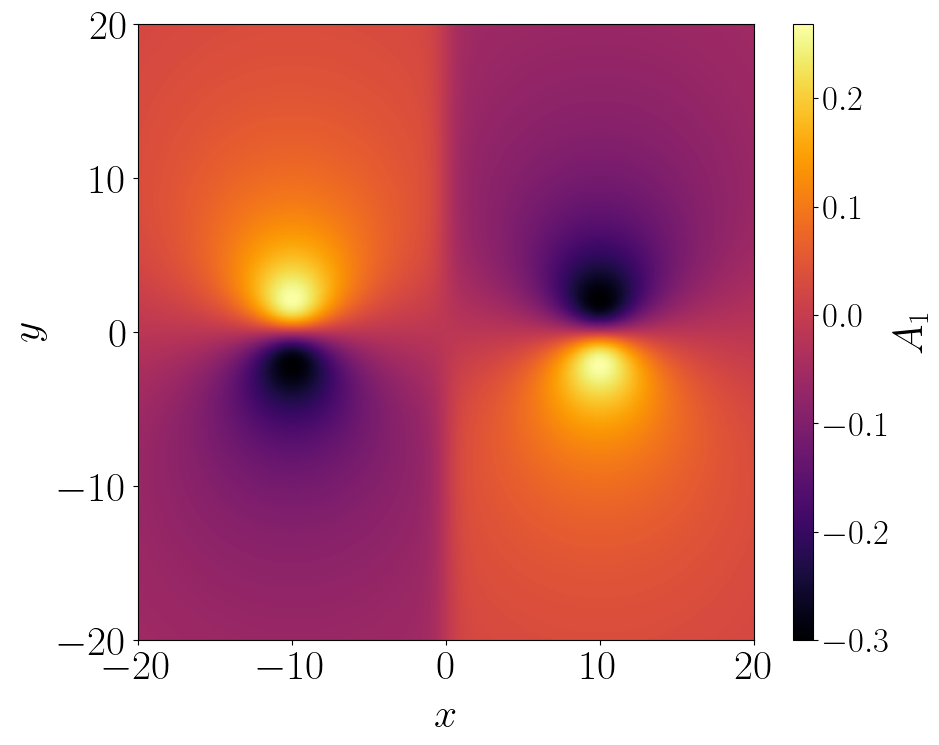}
        \caption{}
    \end{subfigure}
    \hfill
    \begin{subfigure}{0.48\textwidth}
        \includegraphics[width=\linewidth]{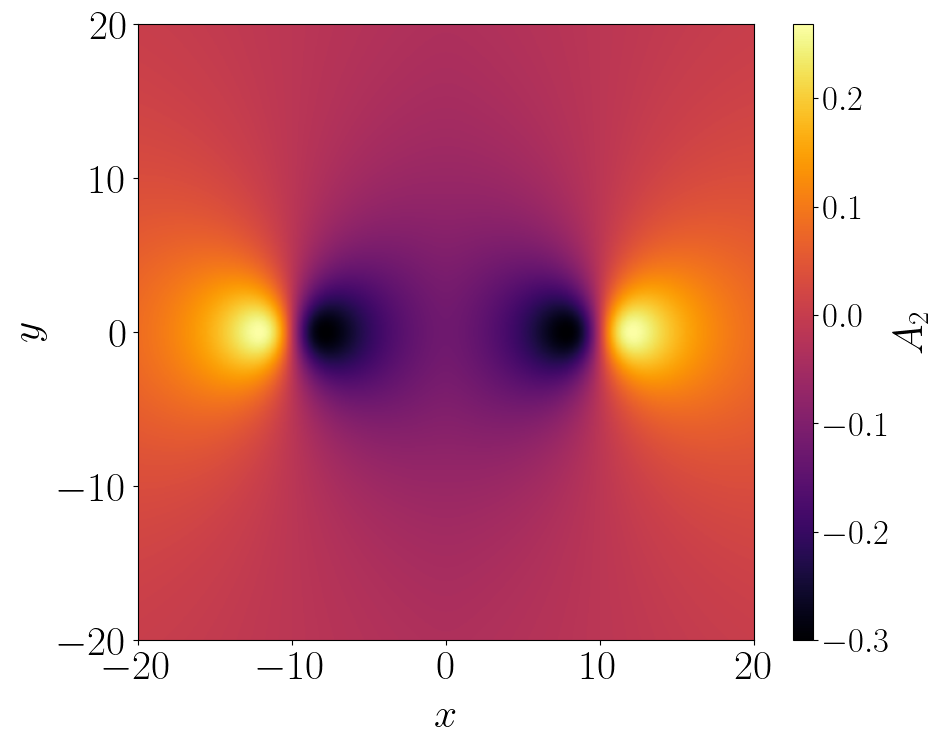}
        \caption{}
    \end{subfigure}
    
    \caption{ Color-maps of the magnetic field (a), Higgs field (b) and gauge components $A_1$ (c) and $A_2$ (d). Parameters are $(a_x, a_y, x_0)=(10, 0, 0)$ and the lattice has size $500 \times 500$ with $L=20$.}

    \label{fig:1}
    
\end{figure}

If one chooses $\sigma_1(x)=\coth (x-x_0)$, the impurity is translated or, equivalently, the center of mass of the av-v pair is shifted. The solitons are then centered at $(\pm a_x+x_0,a_y)$. Thus, the corresponding moduli space is parametrized by the three coordinates $(a_x,a_y,x_0)$, the last two being trivial translational moduli. Figure~\ref{fig:1} shows the fields (\ref{eq:3.14})-(\ref{eq:3.16}) for the choice of parameters $(a_x,a_y,x_0)=(10,0,0)$. Indeed, all solutions represent smooth configurations of a v-av system, except for $A_1$, which changes abruptly across the quadrants of $\mathbb{R}^2$. This behavior originates from the impurity contribution to the $A_1$ gauge-field component, as dictated by eq.~(\ref{eq:3.15}), which becomes significant near the line $x=0$.  We emphasize, however, that $A_1$ is a continuous function and the observed effect is merely a small curvature introduced by the impurity function.

Certainly, the coordinate $a_x$ (or $\alpha_x=\cosh a_x$) controls the collision, since decreasing it brings the two solitons closer together, hence mimicking the usual attractive interaction. This is a rather curious feature. In a conventional two-vortex moduli space, four coordinates are required to specify the vortex positions and describe their collision. By contrast, the BPS structure of the present vortex-impurity model resembles that of kink-impurity systems, where the dynamics is dictated by a single modulus.

\begin{figure}[h!]
    \centering
    % Primeira linha (2 figuras)
    \begin{subfigure}{0.32\textwidth}
        \includegraphics[width=\linewidth]{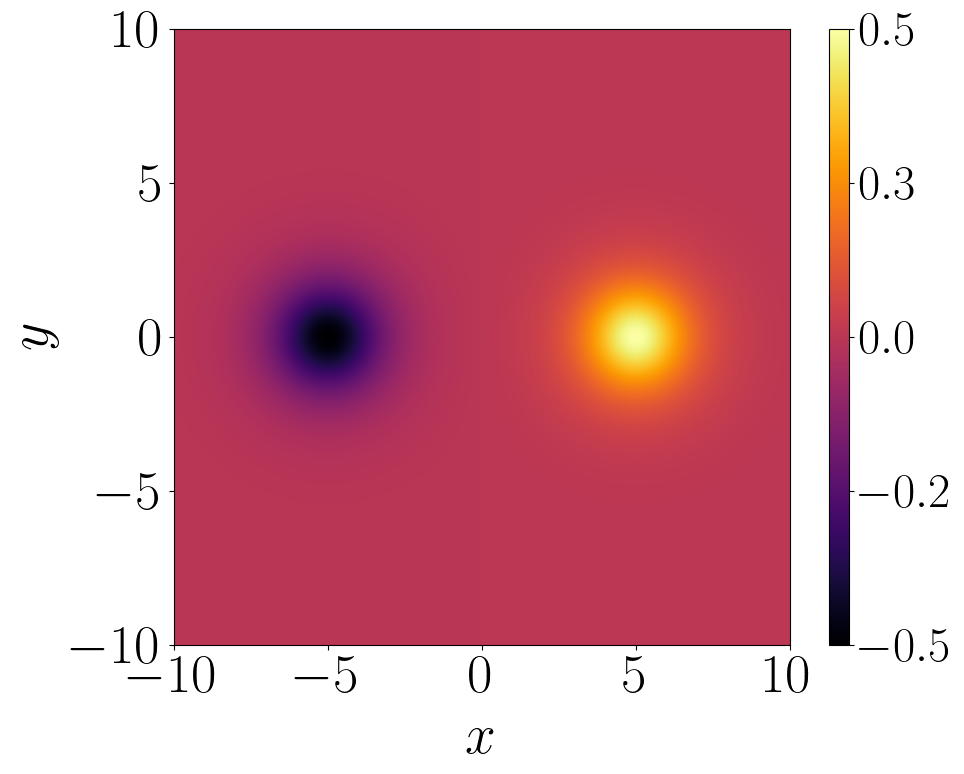}
        \caption{$a_x=5$}
    \end{subfigure}
    \hfill
    \begin{subfigure}{0.32\textwidth}
        \includegraphics[width=\linewidth]{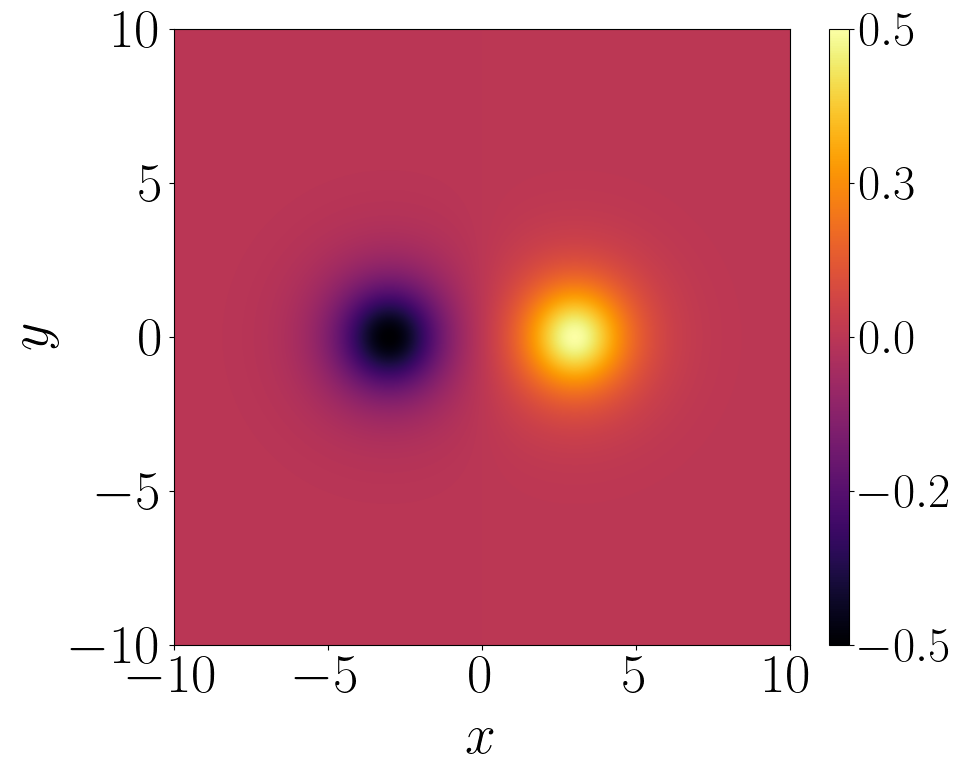}
        \caption{$a_x=3$}
    \end{subfigure}
    \hfill
    \begin{subfigure}{0.32\textwidth}
        \includegraphics[width=\linewidth]{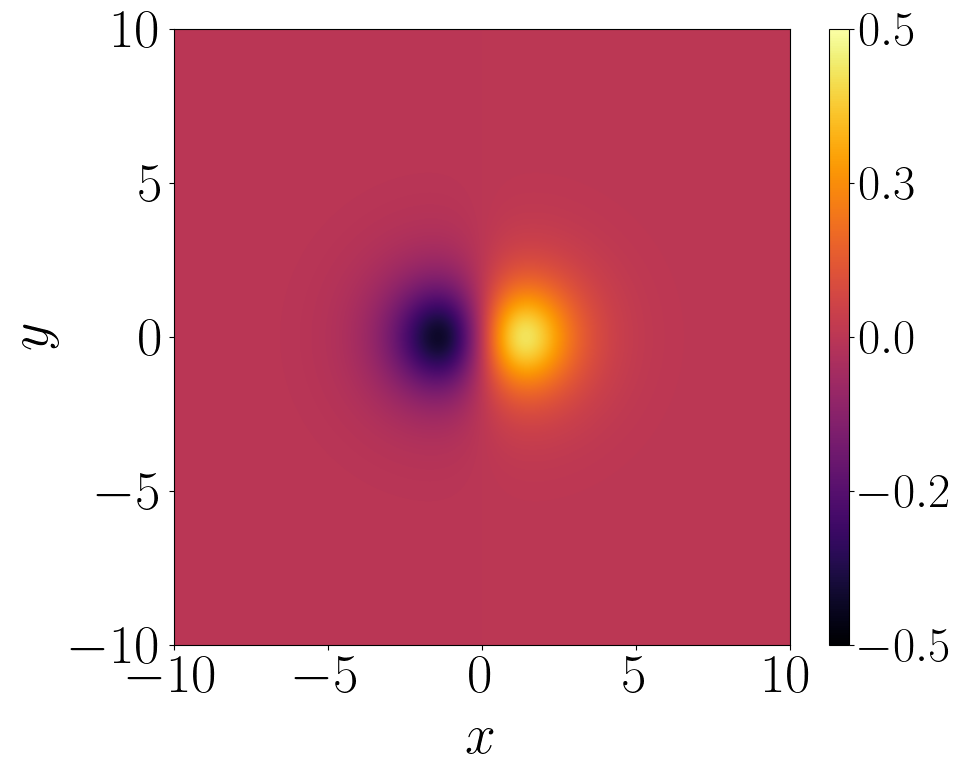}
        \caption{$a_x=1$}
    \end{subfigure}
    
    \vspace{0.01cm} % Espaço vertical entre as linhas
    
    % Segunda linha (2 figuras)
    \begin{subfigure}{0.32\textwidth}
        \includegraphics[width=\linewidth]{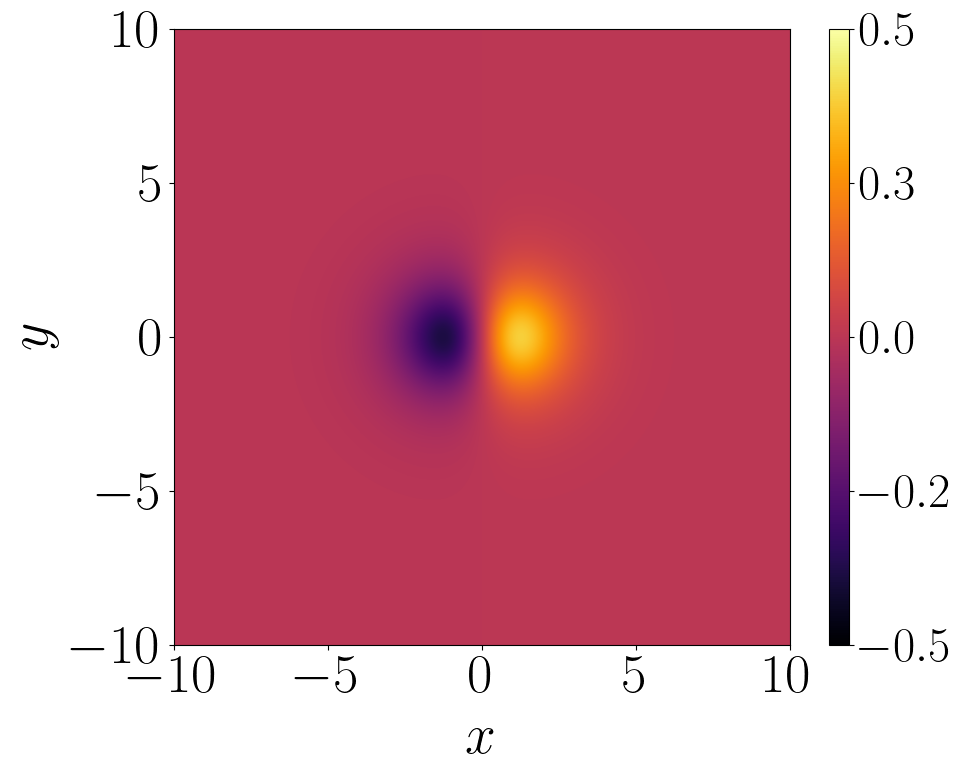}
        \caption{$a_x=0.5$}
    \end{subfigure}
    \hfill
    \begin{subfigure}{0.32\textwidth}
        \includegraphics[width=\linewidth]{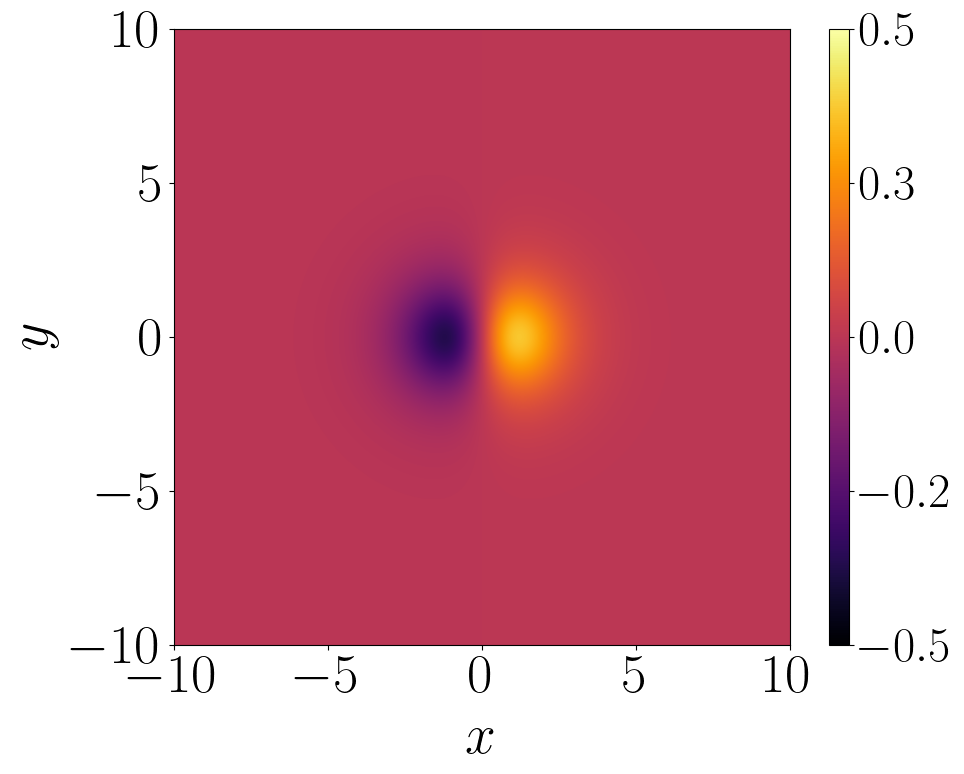}
        \caption{$a_x=0.1$}
    \end{subfigure}
    \hfill
    \begin{subfigure}{0.32\textwidth}
        \includegraphics[width=\linewidth]{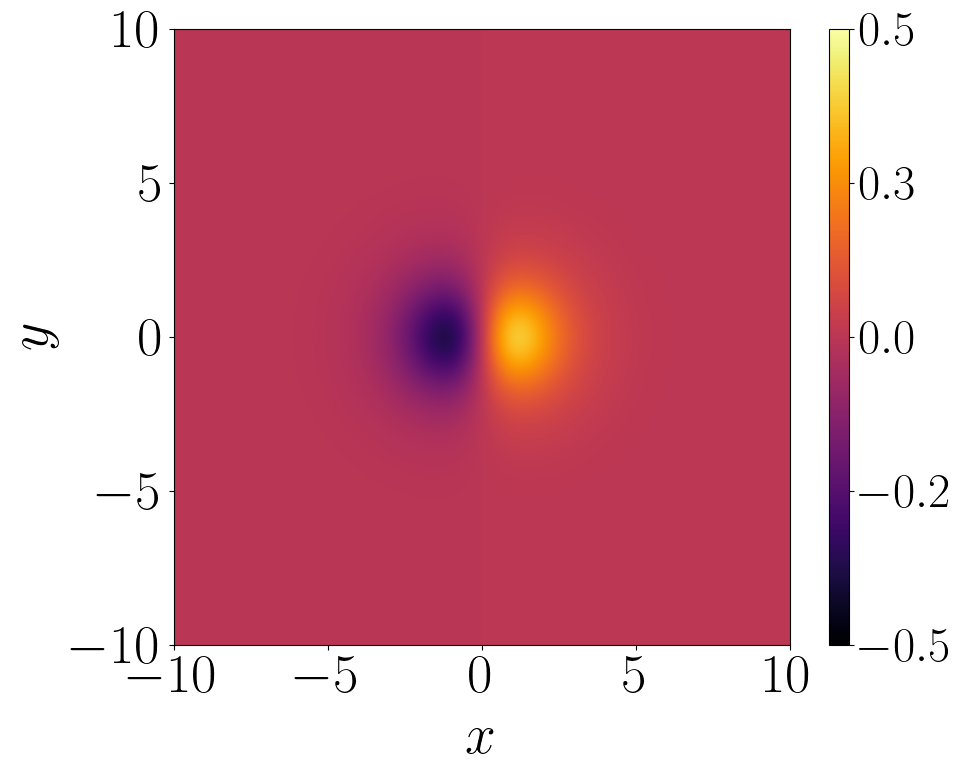}
        \caption{$a_x=0.01$}
    \end{subfigure}

    \vspace{0.01cm} % Espaço vertical entre as linhas
    
    % Segunda linha (2 figuras)
    \begin{subfigure}{0.32\textwidth}
        \includegraphics[width=\linewidth]{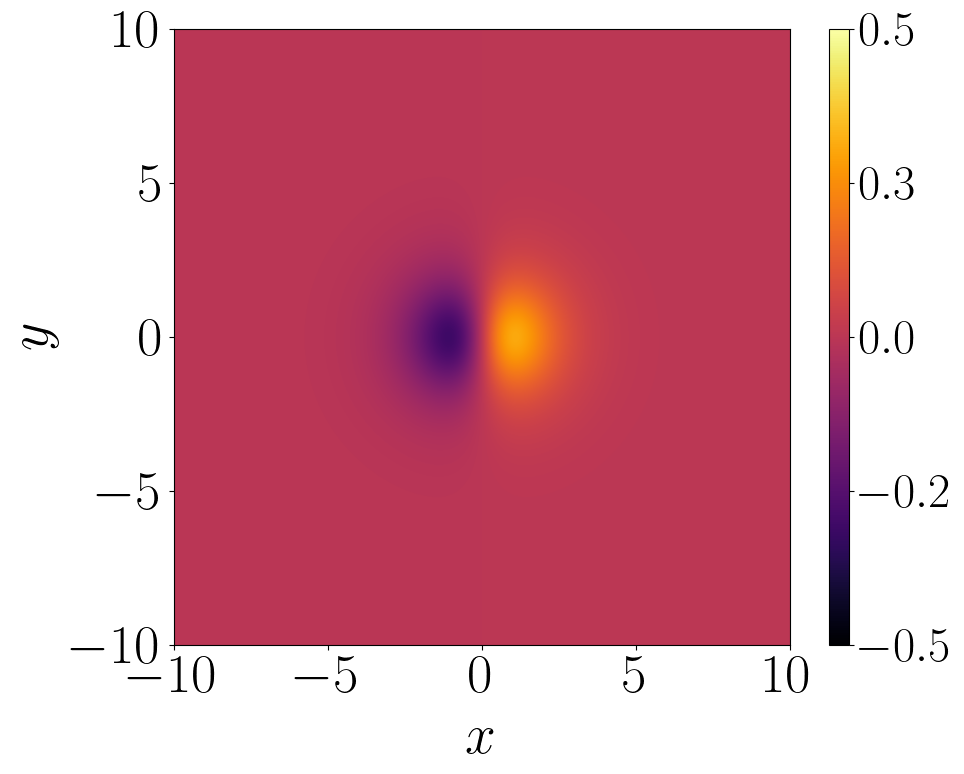}
        \caption{$\alpha_x=0.7$}
    \end{subfigure}
    \hfill
    \begin{subfigure}{0.32\textwidth}
        \includegraphics[width=\linewidth]{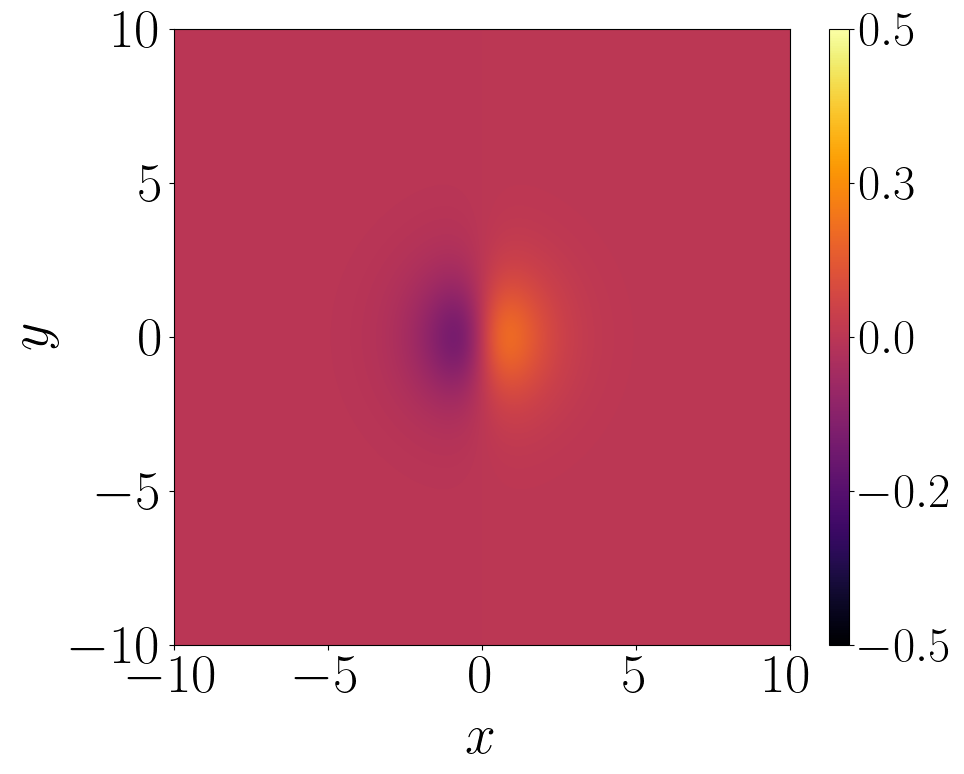}
        \caption{$\alpha_x=0.3$}
    \end{subfigure}
    \hfill
    \begin{subfigure}{0.32\textwidth}
        \includegraphics[width=\linewidth]{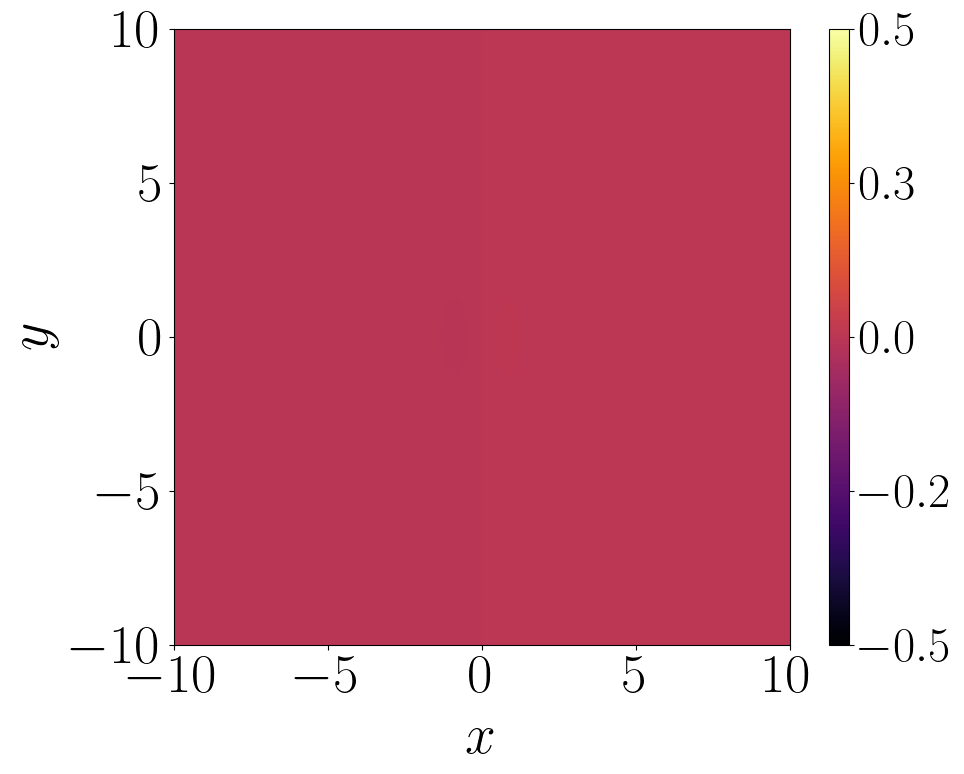}
        \caption{$\alpha_x=0.01$}
    \end{subfigure}

    \caption{Plots of the magnetic field for fixed values of $a_x$.}
    
    \label{fig:2}
\end{figure}

On the other hand, as $a_x\to0$, the solitons approach each other and begin to interact until the configuration reaches a threshold value, $a_{\rm th}\approx0.1$ (figure~\ref{fig:2}(a-f)). Since the solutions are symmetric under the transformation $a_x\to-a_x$, changing the sign of $a_x$ does not alter the field configuration. As discussed earlier, we therefore consider the range $a_x\in[0i,\frac{\pi}{2}i)$, allowing us to use values $0<\alpha_x<1$ and therefore cover the entire moduli space of the mirror impurity. With this choice of branch cut, the collision describes the annihilation of the solitons as $\alpha_x\to0$, as shown in figure~\ref{fig:2}(g-i) for $\alpha_x=0.7$, $0.3$, and $0.01$.

%\enlargethispage{1.5cm}

\begin{figure}[h!]
    \centering
    % Primeira linha (2 figuras)
    \begin{subfigure}{0.48\textwidth}
        \includegraphics[width=\linewidth]{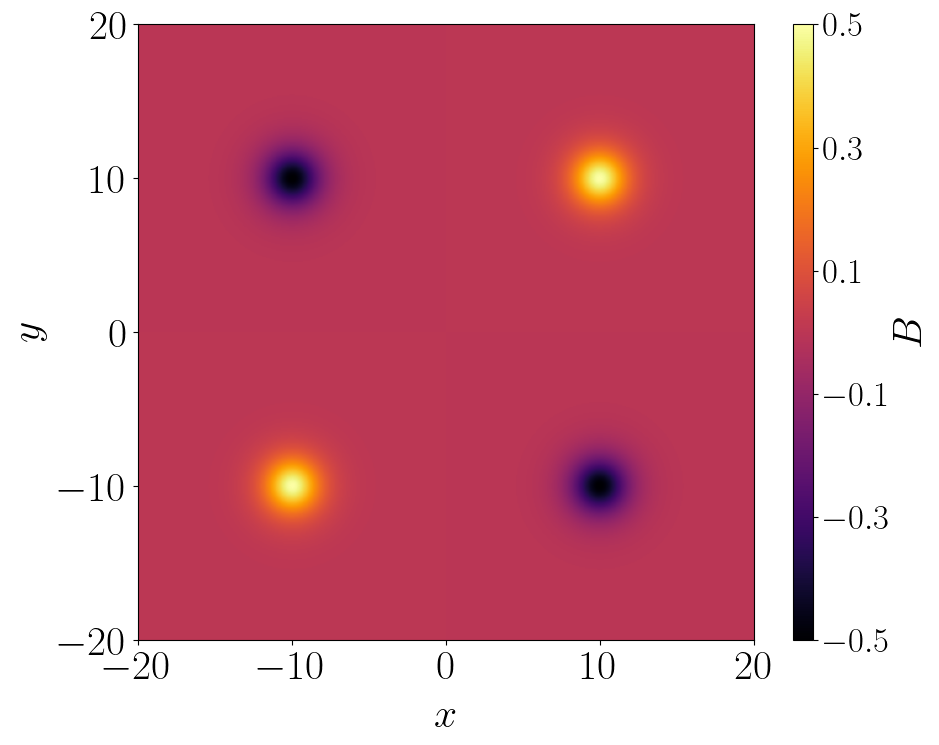}
        \caption{}
    \end{subfigure}
    \hfill
    \begin{subfigure}{0.48\textwidth}
        \includegraphics[width=\linewidth]{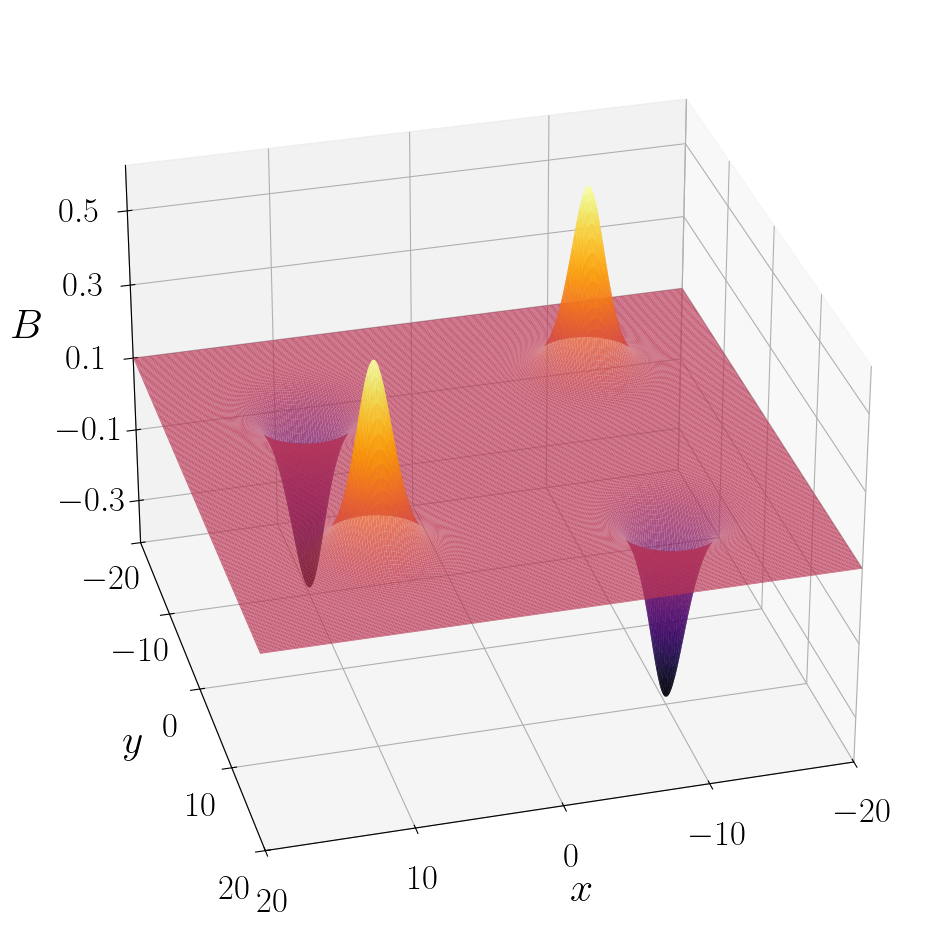}
        \caption{}
    \end{subfigure}

    \vspace{0.01cm}

    \begin{subfigure}{0.48\textwidth}
        \includegraphics[width=\linewidth]{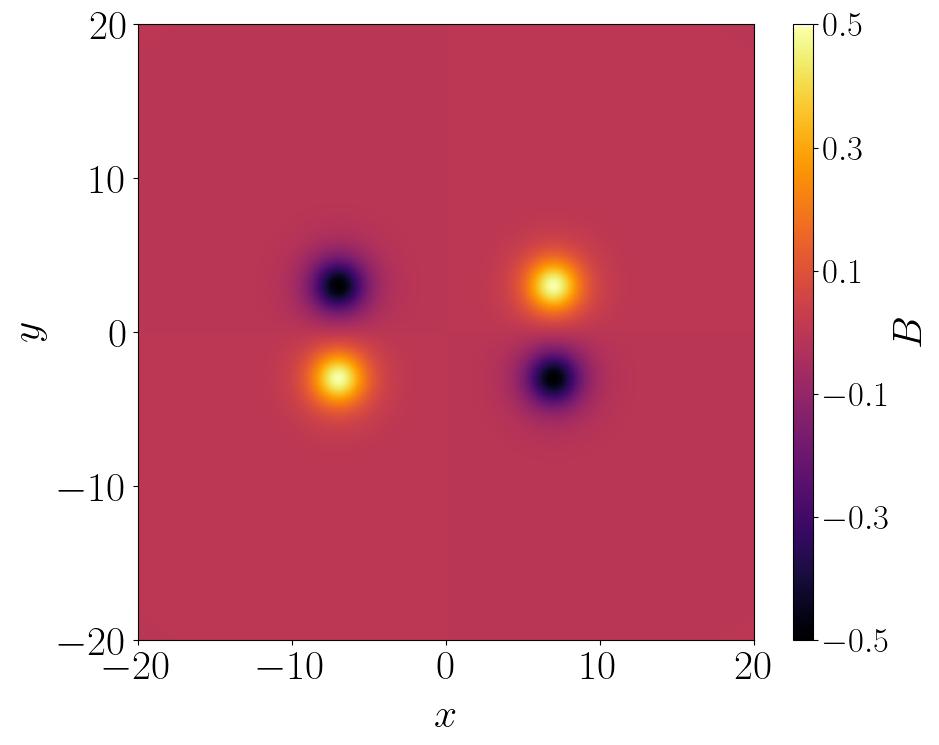}
        \caption{}
    \end{subfigure}
    \hfill
    \begin{subfigure}{0.48\textwidth}
        \includegraphics[width=\linewidth]{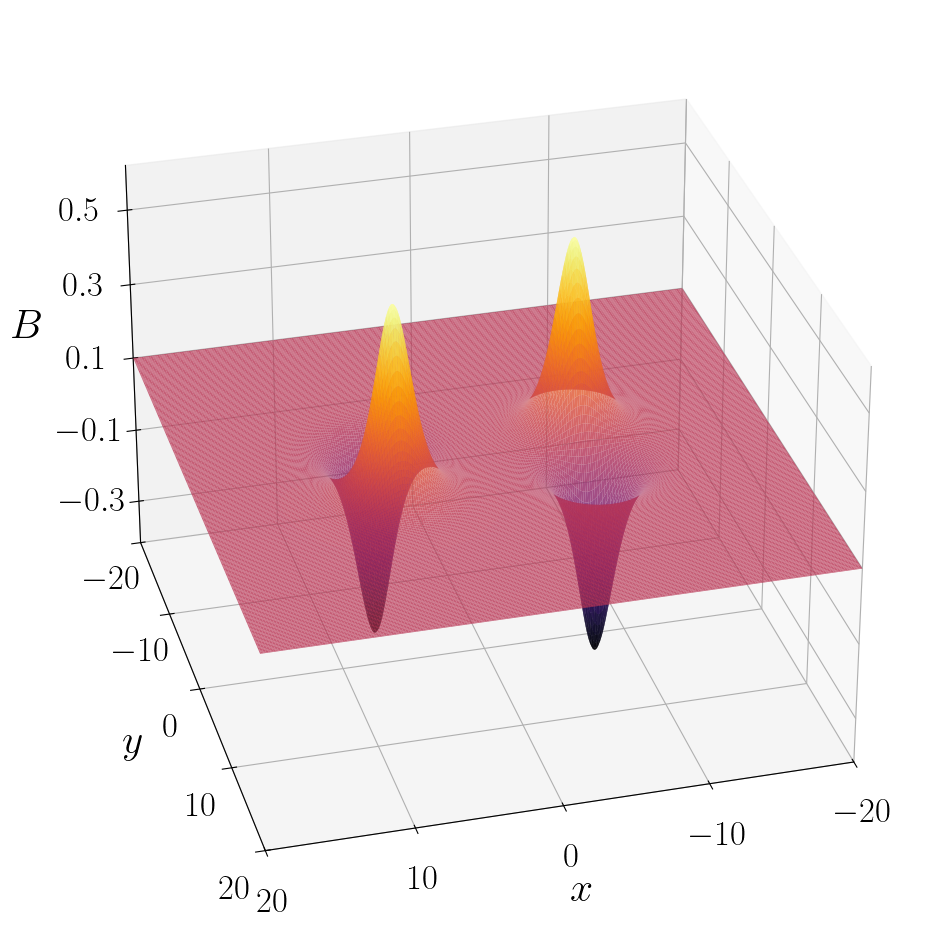}
        \caption{}
    \end{subfigure}
    
    \caption{Representations of the magnetic field solution for the impurities $\sigma_1(x)=\coth(x- x_0)$ and $\sigma_2(y)=\coth(y-y_0)$. Parameters are $(a_x, a_y, x_0, y_0)=(10, 0, 0, 0)$ (a-b) and $(a_x, a_y, x_0, y_0)=(7, 3, 0, 0)$ (c-d).}

    \label{fig:3}
\end{figure}

So far, the chosen impurity has acted as a mirror placed at $x=x_0$. We now apply the same construction by setting $\sigma_2(y)=\coth(y-y_0)$, which reflects the previous solutions across the line $y=y_0$. The resulting configuration consists of a rectangular arrangement of vortices and antivortices, with identical solitons occupying opposite vertices: vortices centered at $(\pm a_x+x_0,\pm a_y+y_0)$ and antivortices at $(\mp a_x+x_0,\pm a_y+y_0)$ (figure~\ref{fig:3}). In this sense, coupling the two impurities is equivalent to placing two perpendicular mirrors.

The addition of the impurity $\sigma_2(y)=\coth(y-y_0)$ introduces a new modulus, $y_0$, which shifts the center of mass of the lattice along the $y$-axis. Likewise, the coordinate $a_y$ plays the same role as $a_x$, controlling the separation between the solitons in the vertical direction. Finally, this construction introduces two additional constraints, analogous to eq.~(\ref{eq:3.19}), along the line $y=y_0$.

Interestingly, this coupling of impurities can be used to generalize the construction to more intricate arrangements of vortices and antivortices. In particular, the next section presents an iterative scheme, inspired by the method introduced by Manton for kink-antikink systems \cite{manton2019}, to generate such configurations.

\newpage

\subsection{Iterative scheme} 
\label{sec:3.B}

The iterative scheme proposed here extends the mirror-impurity construction to the solutions of the equations
\begin{align}
&\sigma_1^{(n_1)}(x) D_1 \phi + i\sigma_2^{(n_2)}(y) D_2 \phi = 0 \;\; \label{eq:3.20}\text{and} \\
&\sigma_1^{(n_1)}(x)\sigma_2^{(n_2)}(y) B + \frac{1}{2}(|\phi|^2 - 1) = 0 , \;\;\;n_1, n_2 = 0, 1, 2, ... \; ,\label{eq:3.21}
\end{align} 
where, to simplify the notation, we suppressd the indices $(n_1,n_2)$ on the fields, writing $\phi\equiv\phi^{(n_1,n_2)}$ and $B\equiv B^{(n_1,n_2)}$. We begin with $\sigma_1^{(0)}=\sigma_2^{(0)}=1$, and generate the subsequent iterations according to
\begin{align}
&\sigma_1^{(n_1)}(x) = \coth \left(\int_{a^{(n_1-1)}_x}^x \frac{dx'}{\sigma_1^{(n_1-1)}(x')}\right) ,\label{(eq:3.22)} \\
&\sigma_2^{(n_2)}(y) = \coth \left(\int_{a^{(n_2-1)}_y}^y \frac{dy'}{\sigma_2^{(n_2-1)}(y')}\right), \label{eq:3.23}
\end{align} 
where $a_i^{(n_i)}$ $(i=1,2)$ denote the moduli introduced at each iteration. In other words, the $n_1$-th (or $n_2$-th) impurity is obtained by applying the $\coth$ function to the coordinate $u^{(n_1-1)}$ (or $v^{(n_2-1)}$) generated in the previous iteration.

Within this framework, the zeroth iteration corresponds to a single vortex centered at the origin\footnote{We neglect the translational moduli, $a_x^{(0)}$ and $a_y^{(0)}$, since they do not introduce any nontrivial dynamics into the theory.}. The independent indices $n_1$ and $n_2$ allow the iterative procedure to be carried out separately along the two coordinate directions, creating different v-av arrangements. As a first illustration of the method, we perform successive iterations only along the $x$ direction, fixing $\sigma_2^{(n_2)}=\sigma_2^{(0)}$. This generates a horizontal chain of alternating vortices and antivortices, as we show below.

The first iteration is obtained by choosing $\sigma_1^{(1)} = \coth(x)$ and the solution is the av-v pair we presented in section \ref{sec:3.A}, labeled by the modulus $a_x^{(1)}$ (or $\alpha_x = \cosh a_x^{(1)}$). The next family of solutions is generated for $n_1=2$, with the impurity
\begin{align}
&\sigma_1^{(2)}(x) = \coth \left[\ln \left(\frac{\cosh x}{\cosh a_x^{(1)}} \right) \right] = \frac{\cosh^2 x + \alpha_x^2}{\cosh^2 x - \alpha_x^2} , \label{eq:3.24}
\end{align}
while the corresponding coordinate becomes
\begin{align}
&u^{(2)}(x) =  x - \beta_x - 2\frac{\alpha_x}{\sqrt{1 + \alpha_x^2}}\tanh^{-1} \left(\frac{\alpha_x}{\sqrt{1 + \alpha_x^2}} \tanh x \right) , \label{eq:3.25}
\end{align} 
where $\beta_x$ is a function of the moduli $a_x^{(2)}$ and $\alpha_x$. 

\begin{figure}[h!]
    \centering
    % Primeira linha (2 figuras)
    \begin{subfigure}{0.32\textwidth}
        \includegraphics[width=\linewidth]{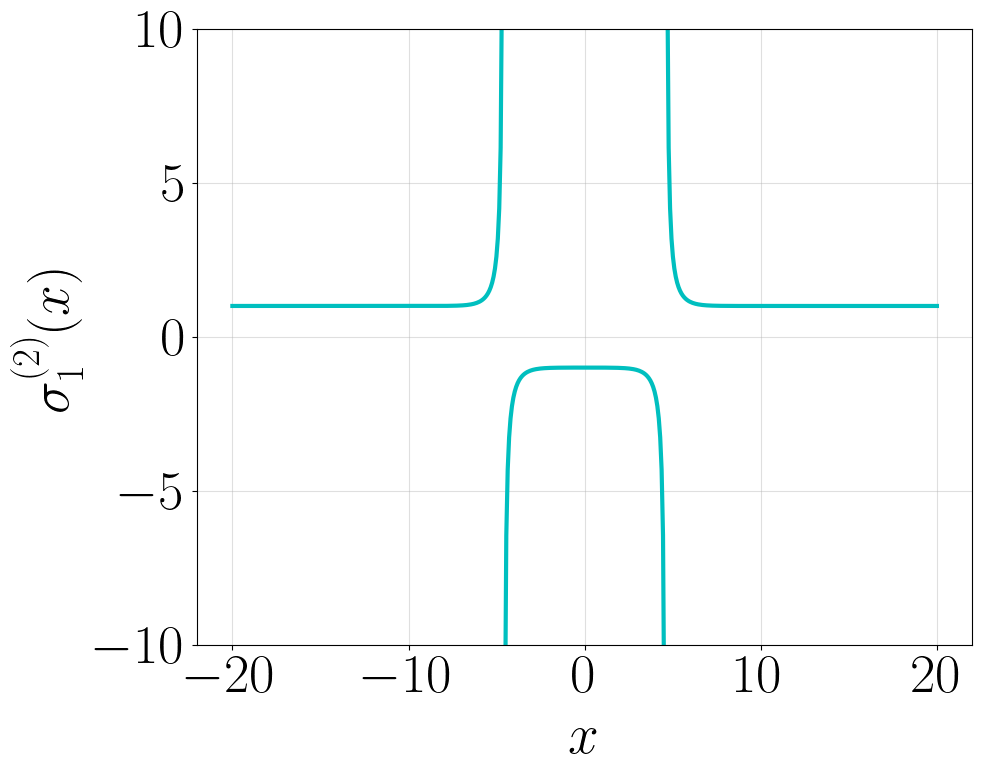}
        \caption{}
    \end{subfigure}
    \hfill
    \begin{subfigure}{0.32\textwidth}
        \includegraphics[width=\linewidth]{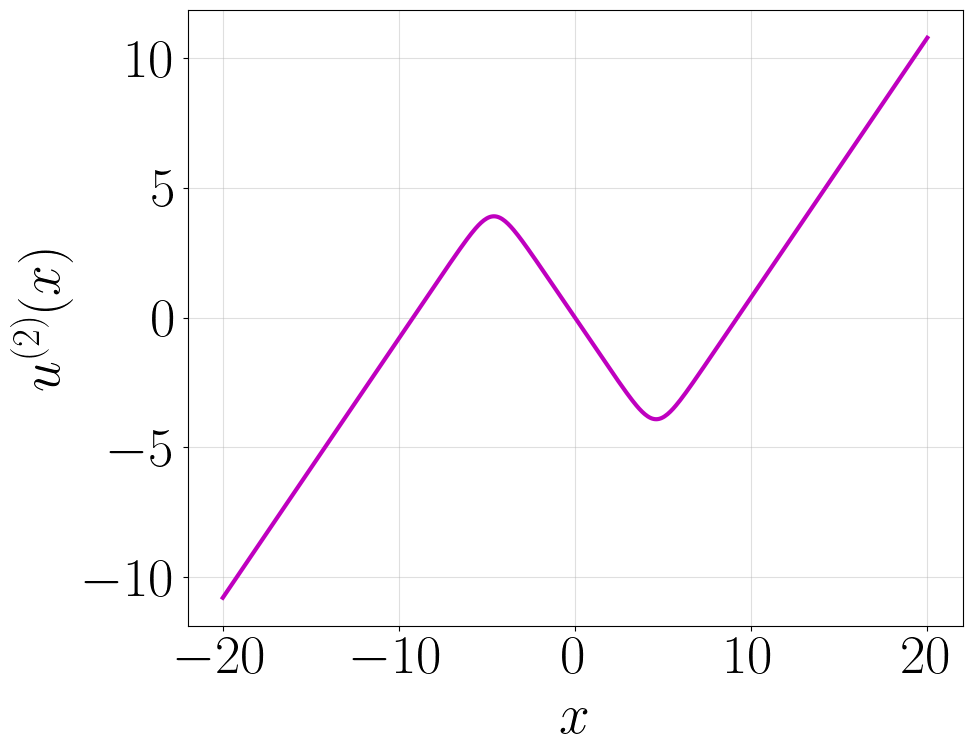}
        \caption{}
    \end{subfigure}
    \hfill
    \begin{subfigure}{0.32\textwidth}
        \includegraphics[width=\linewidth]{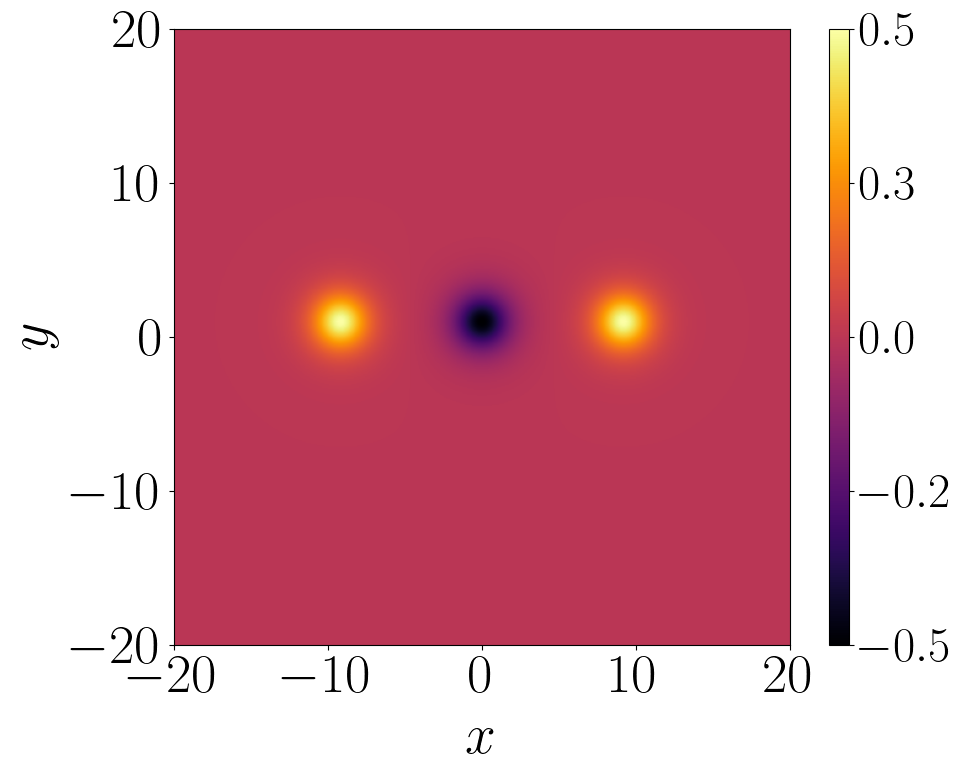}
        \caption{$\alpha_x = 50$}
    \end{subfigure}

    \vspace{0.1cm} % Espaço vertical entre as linhas
    
    % Segunda linha (2 figuras)
    \begin{subfigure}{0.32\textwidth}
        \includegraphics[width=\linewidth]{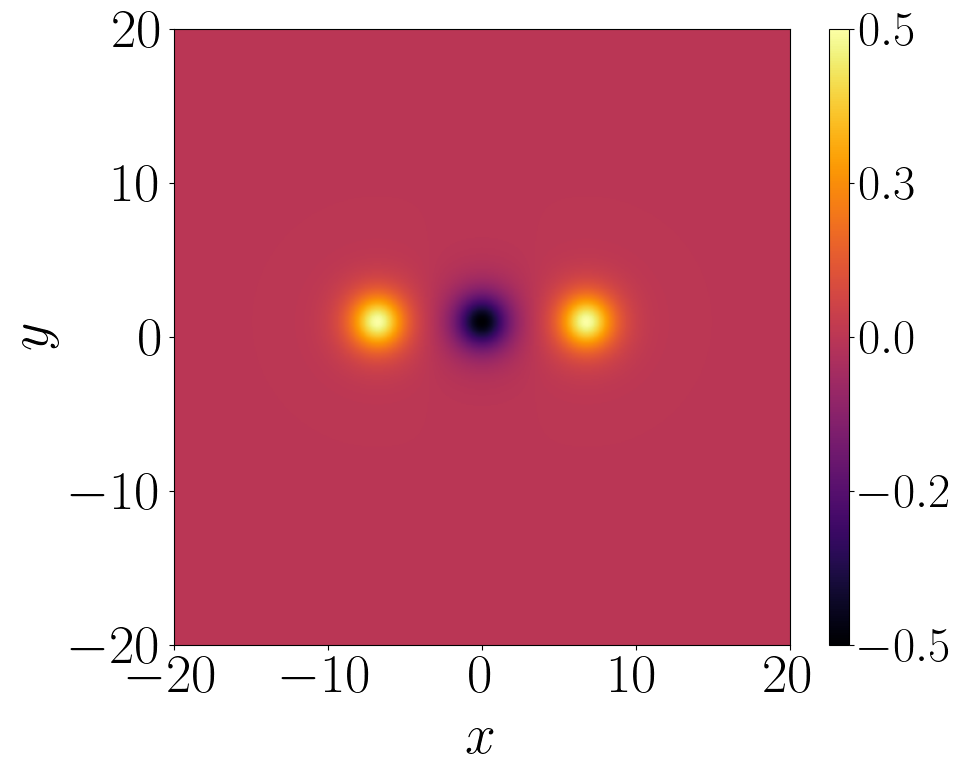}
        \caption{$\alpha_x = 15$}
    \end{subfigure}
    \hfill
    \begin{subfigure}{0.32\textwidth}
        \includegraphics[width=\linewidth]{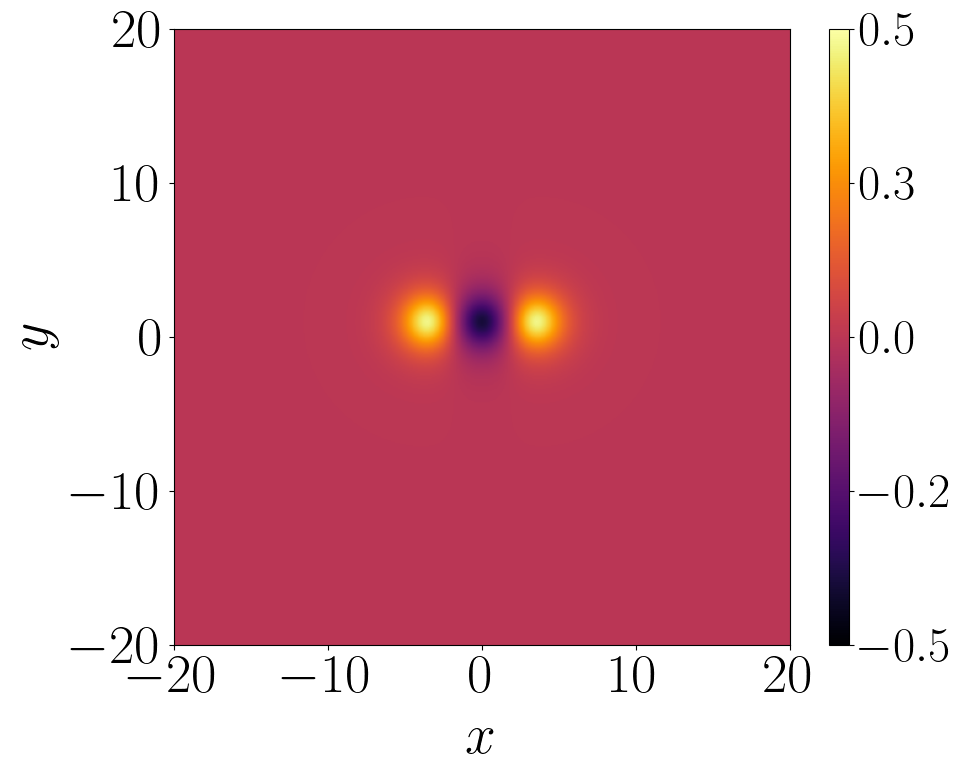}
        \caption{$\alpha_x = 3$}
    \end{subfigure}
    \hfill
    \begin{subfigure}{0.32\textwidth}
        \includegraphics[width=\linewidth]{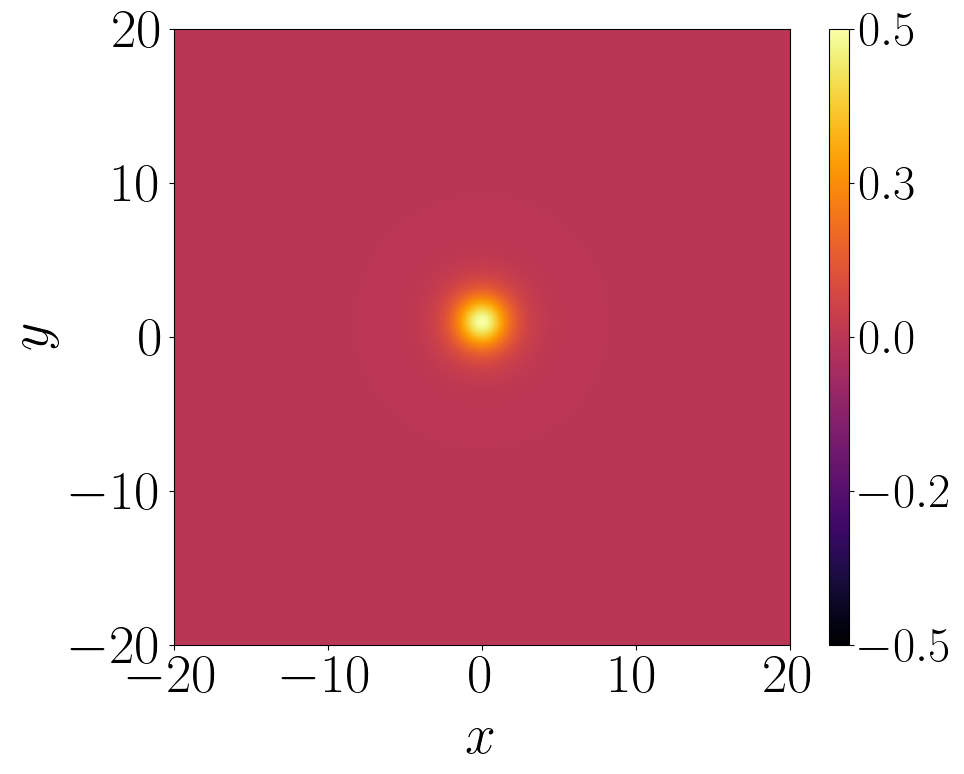}
        \caption{$\alpha_x=0.01$}
    \end{subfigure}
    
    \caption{(a-b): Impurity and coordinate for the second iteration ($\alpha_x=50$, $\beta_x=0$). (c): Magnetic field $B^{(2,0)}$ with the same parameters. (d-f): $B^{(2,0)}$ for several values of $\alpha_x$, fixing $\beta_x=0$.}
    
    \label{fig:4}
\end{figure}

Figures~\ref{fig:4}(a) and \ref{fig:4}(b) show the impurity (\ref{eq:3.24}) and the corresponding coordinate (\ref{eq:3.25}), respectively, for $\alpha_x=50$ and $\beta_x=0$. One can interpret $\sigma_1^{(2)}$ as a combination of two mirror-impurities. The first, centered at $(-a_x^{(1)},0)$, generates a v-av pair, with the impurity located at their midpoint, as expected. The second mirror impurity, centered at $(a_x^{(1)},0)$, reflects the central antivortex into a second vortex. As before, the soliton centers are determined by the roots of the equation $u^{(2)}(x)=0$, provided they are sufficiently well separated. These roots cannot be obtained analytically in closed form. However, one can verify their properties by comparing figure \ref{fig:4}(b) with the color map of the corresponding magnetic field solution, $B^{(2, 0)}$, provided in figure \ref{fig:4}(c). 

In addition, decreasing $\alpha_x$ brings the two mirror impurities closer together or, equivalently, moves the vortices toward the central antivortex (figure~\ref{fig:4}(c-e)). Throughout this process, the vortex-antivortex-vortex configuration is preserved until $\alpha_x=1$ ($a_x^{(1)}=0$), where the centers of the mirror impurities overlap. As a result, the central antivortex disappears while the two vortices merge, leading to a single-vortex configuration in the limit $\alpha_x\to0$ (figure~\ref{fig:4}(f)).

\begin{figure}[b!]
    \centering
    % Primeira linha (2 figuras)
    \begin{subfigure}{0.32\textwidth}
        \includegraphics[width=\linewidth]{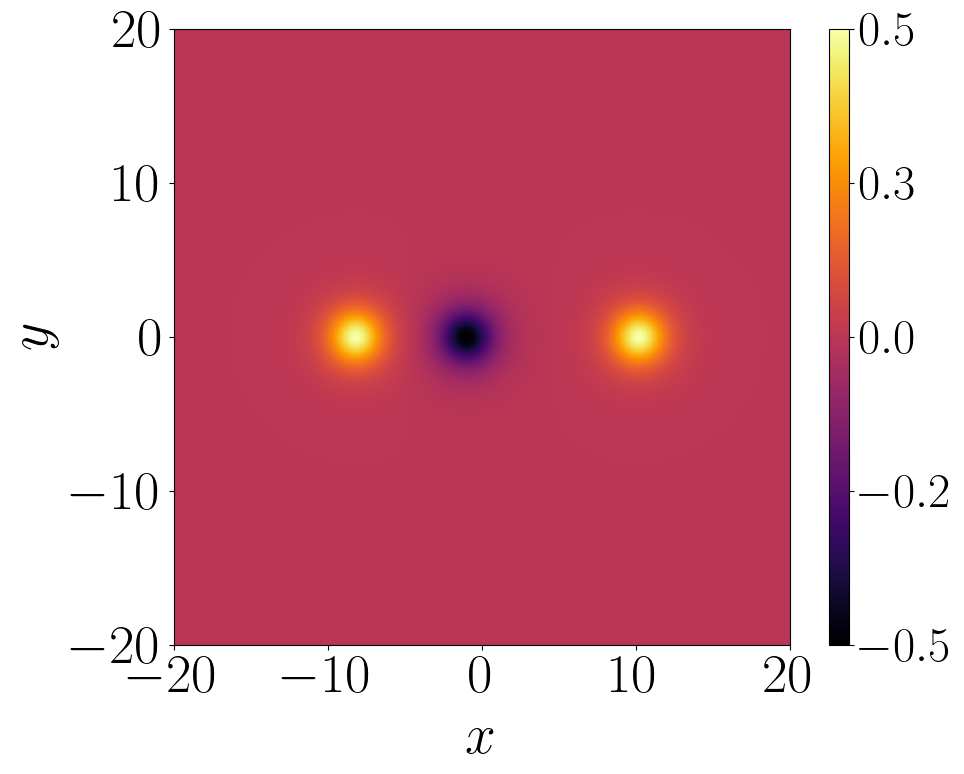}
        \caption{$\beta_x=1$}
    \end{subfigure}
    \hfill
    \begin{subfigure}{0.32\textwidth}
        \includegraphics[width=\linewidth]{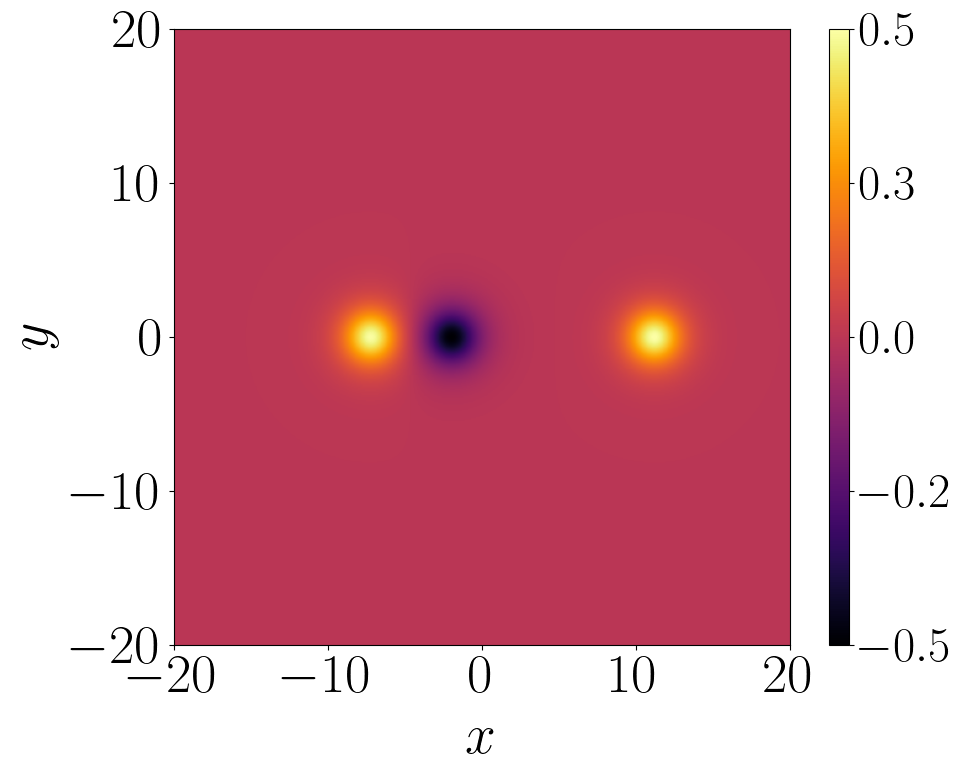}
        \caption{$\beta_x=2$}
    \end{subfigure}
    \hfill
    \begin{subfigure}{0.32\textwidth}
        \includegraphics[width=\linewidth]{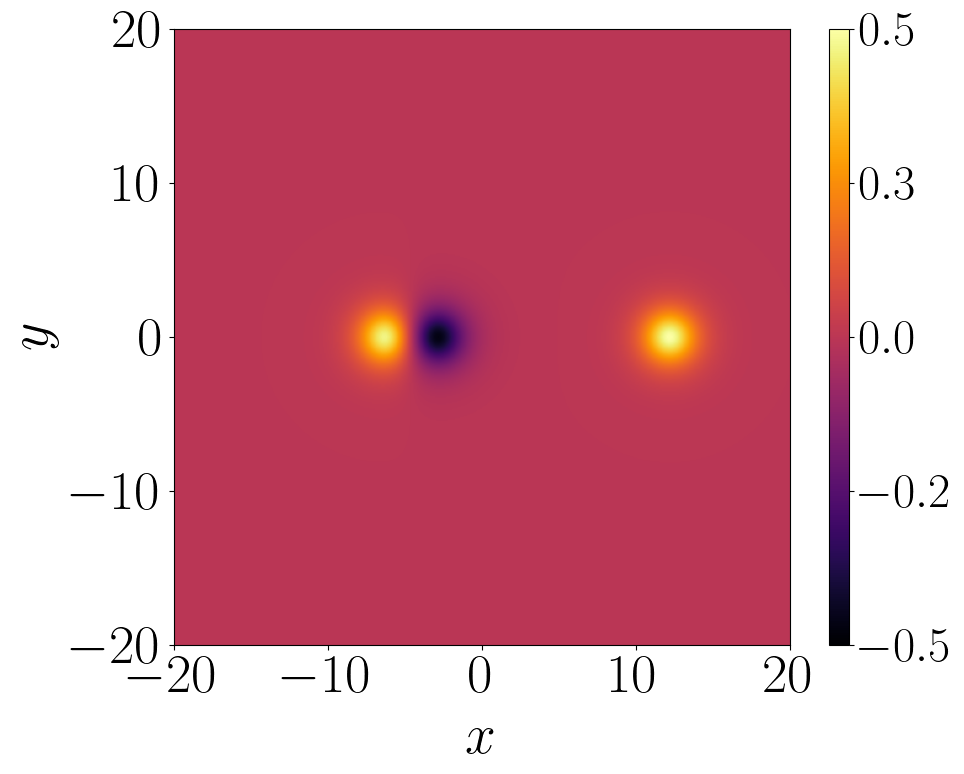}
        \caption{$\beta_x=3$}
    \end{subfigure}
    
    \vspace{0.01cm} % Espaço vertical entre as linhas
    
    % Segunda linha (2 figuras)
    \begin{subfigure}{0.32\textwidth}
        \includegraphics[width=\linewidth]{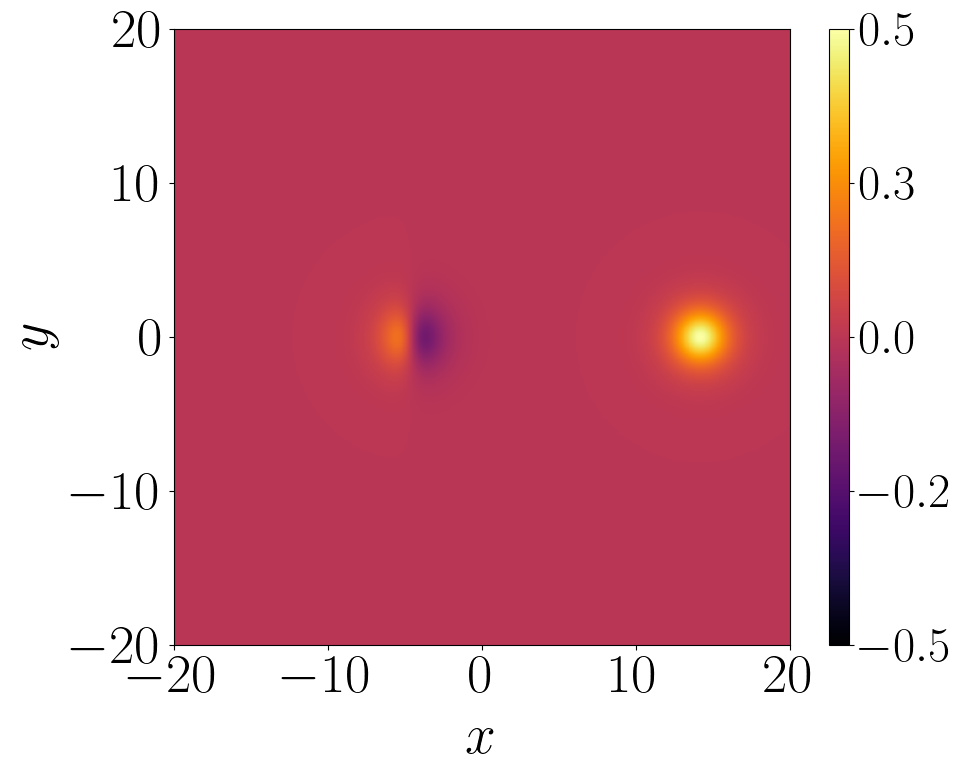}
        \caption{$\beta_x=5$}
    \end{subfigure}
    \hfill
    \begin{subfigure}{0.32\textwidth}
        \includegraphics[width=\linewidth]{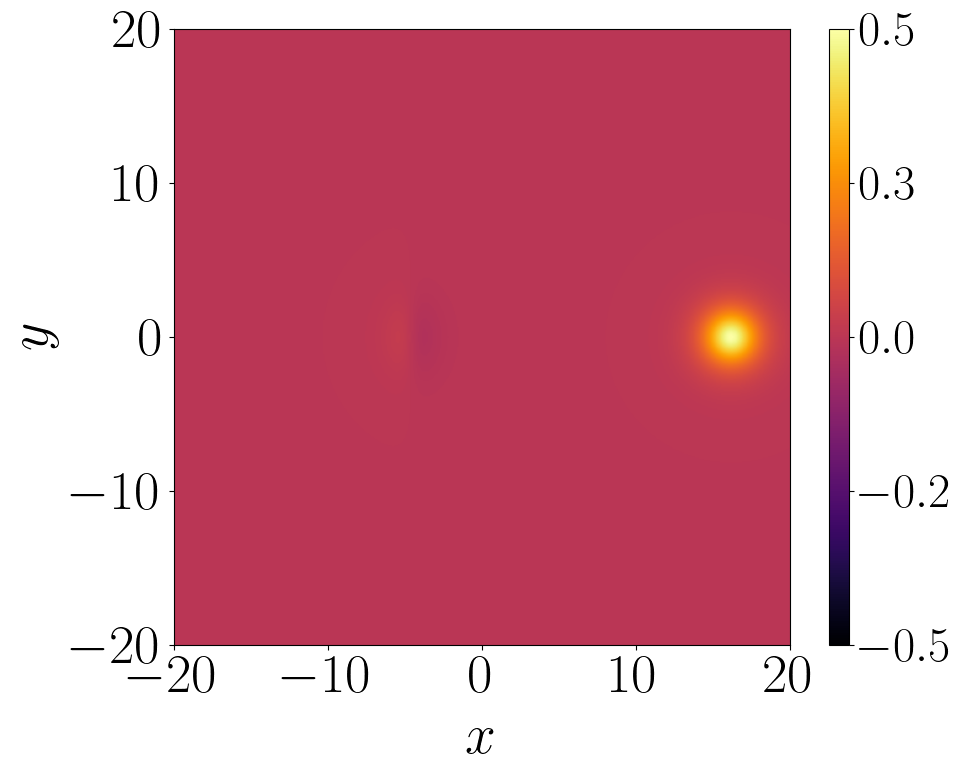}
        \caption{$\beta_x=7$}
    \end{subfigure}
    \hfill
    \begin{subfigure}{0.32\textwidth}
        \includegraphics[width=\linewidth]{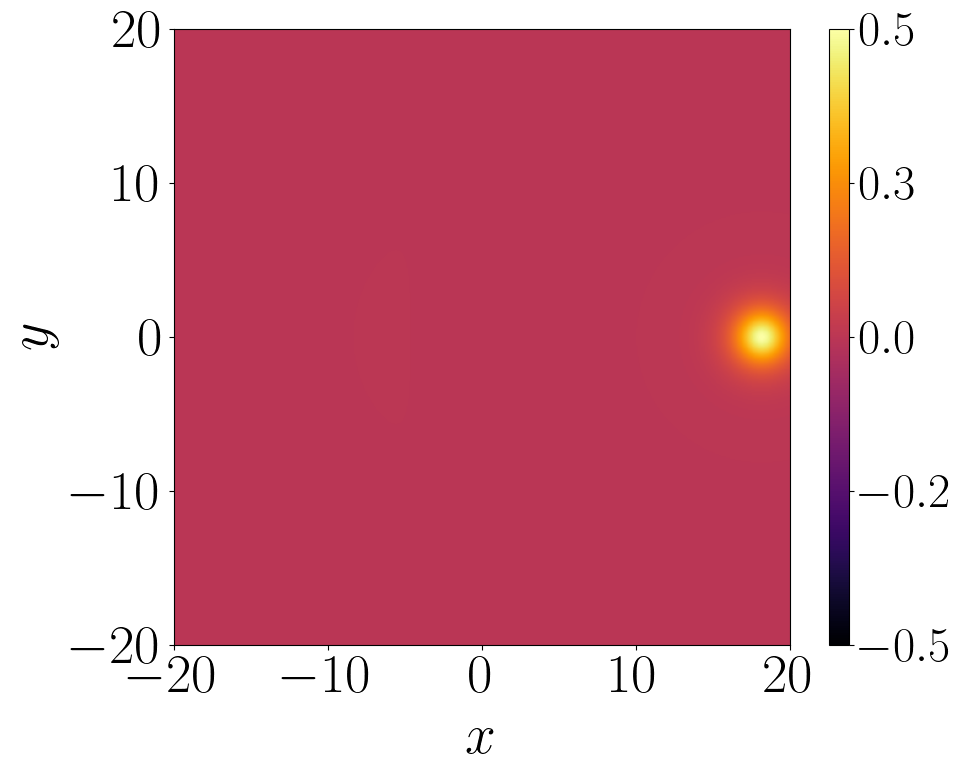}
        \caption{$\beta_x=9$}
    \end{subfigure}
    
    \caption{Plots of the magnetic field $B^{(2, 0)}$ for several values of $\beta_x$. We fix $\alpha_x=50$.}
    
    \label{fig:5}
\end{figure}

On the other hand, varying $\beta_x$ translates the function shown in figure~\ref{fig:4}(b), possibly changing the number of roots of the coordinate $u^{(2)}$ and, consequently, the number of solitons. For $\beta_x>0$, increasing its value brings the two negative roots of $u^{(2)}$ closer together until they coincide at the threshold value $\beta_x=\beta_1$. Thus, for $\beta_x\geq\beta_1$, the v-av pair on the left gradually deforms, evolving into complete annihilation in the limit $\beta_x\gg\beta_1$, while the vortex located at $x>0$ moves away from the pair (figure~\ref{fig:5}). Likewise, decreasing $\beta_x$ below zero produces a similar process on the opposite side, with the av-v pair on the right annihilating when $\beta_x \ll \beta_2$.

For fixed $\alpha_x>1$, the values of $\beta_1$ and $\beta_2$ correspond to the local maximum and minimum, respectively, of the function $u^{(2)}(x)$ evaluated at $\beta_x=0$. These extrema occur at $x=\pm a_x^{(1)}$. Then, it is straightforward to write $\beta_1$ and $\beta_2$ as
\begin{align}
&\beta_1=  a_x^{(1)} - 2\frac{\alpha_x}{\sqrt{1 + \alpha_x^2}}\tanh^{-1} \left(\frac{\alpha_x}{\sqrt{1 + \alpha_x^2}} \tanh a_x^{(1)} \right) = -\beta_2 , \label{eq:3.26}
\end{align}
For $\alpha_x=50$, numerical calculations give $\beta_1=-\beta_2\approx3.9$. Figure~\ref{fig:5} shows the evolution of the collision in moduli space for several values of $\beta_x$, fixing $\alpha_x=50$. As expected, for $\beta_x>\beta_1$ (figure~\ref{fig:5}(d-f)), the v-av pair on the left gradually deforms, eventually disappearing through annihilation.

To summarize, the first iteration gives the av-v solutions discussed in section~\ref{sec:3.A}. For $(n_1,n_2)=(2,0)$, the solutions are parametrized by the moduli $(\alpha_x,\beta_x)$ and describe a smooth v-av-v configuration for $\beta_2<\beta_x<\beta_1$ and $\alpha_x>1$, where $\beta_1$ and $\beta_2$ are given by eq.~(\ref{eq:3.26}). Moreover, for $\beta_x\geq\beta_1$ (or $\beta_x\leq\beta_2$), the collision leads to the annihilation of a v-av pair, leaving behind a single vortex. Finally, as $\alpha_x\to0$, the solutions converge to a single vortex whose position is controlled by the modulus $\beta_x$. The iterative procedure can be continued indefinitely, introducing a new modulus at each step. However, for $n_1>2$, the coordinate $u^{(n_1)}(x)$ no longer admits a closed analytical expression, which increases the numerical error to the solution.

We emphasize that the iterative scheme consists of two independent sequences, $\sigma_1^{(n_1)}$ and $\sigma_2^{(n_2)}$. For example, performing a single iteration for each impurity leads to the solution $B^{(1,1)}$, corresponding to the rectangular lattice discussed in section~\ref{sec:3.A} (figure~\ref{fig:3}), parametrized by the moduli $a_x^{(1)}$ and $a_y^{(1)}$. Likewise, the solution for $(n_1,n_2)=(2,1)$ is obtained by solving eqs.~(\ref{eq:3.20})-(\ref{eq:3.21}) using the impurity $\sigma_1^{(2)}$ given by eq.~(\ref{eq:3.24}) together with $\sigma_2^{(1)}=\coth(y)$, and so forth.

\begin{figure}[h!]
    \centering
    % Primeira linha (2 figuras)
    \begin{subfigure}{0.48\textwidth}
        \includegraphics[width=\linewidth]{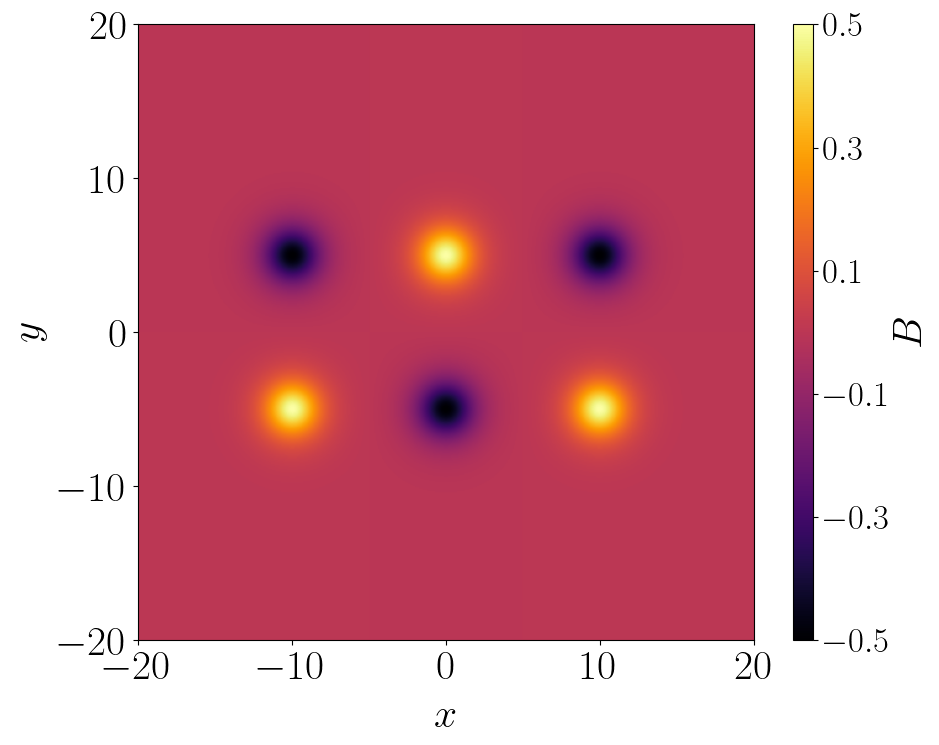}
        \caption{}
    \end{subfigure}
    \hfill
    \begin{subfigure}{0.48\textwidth}
        \includegraphics[width=\linewidth]{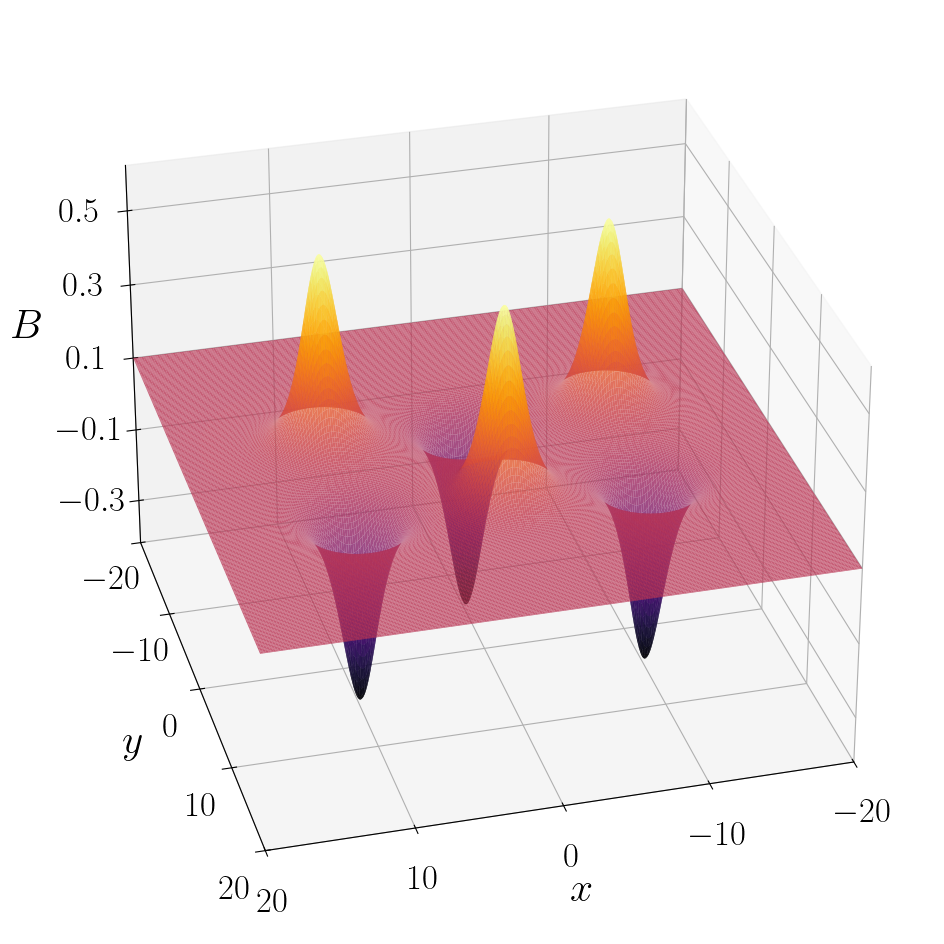}
        \caption{}
    \end{subfigure}

    \vspace{0.01cm} % Espaço vertical entre as linhas

        % Primeira linha (2 figuras)
    \begin{subfigure}{0.48\textwidth}
        \includegraphics[width=\linewidth]{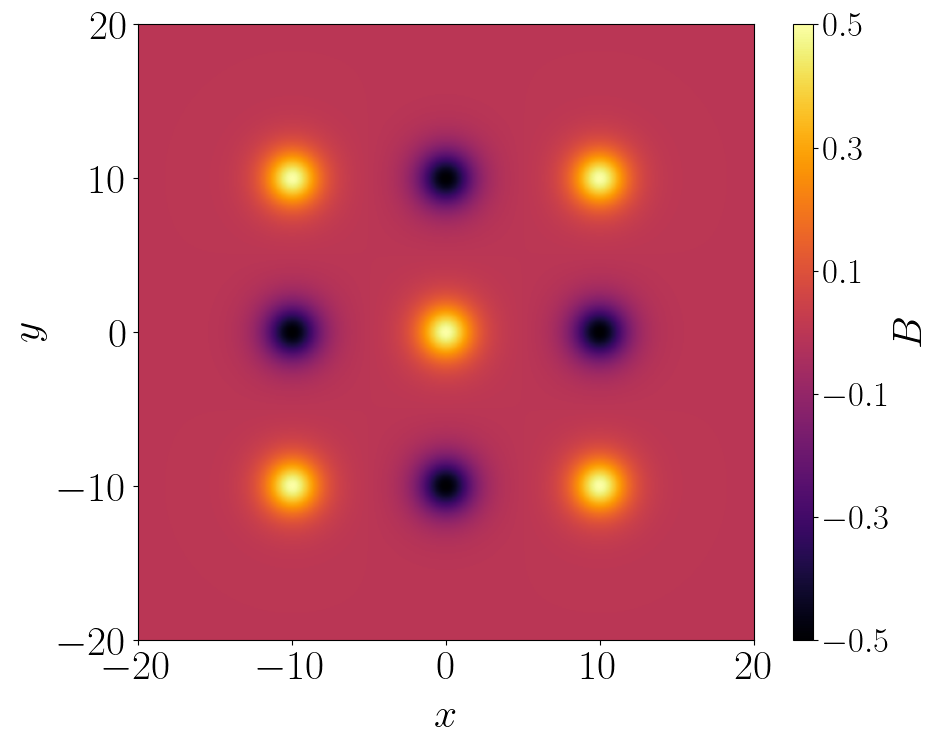}
        \caption{}
    \end{subfigure}
    \hfill
    \begin{subfigure}{0.48\textwidth}
        \includegraphics[width=\linewidth]{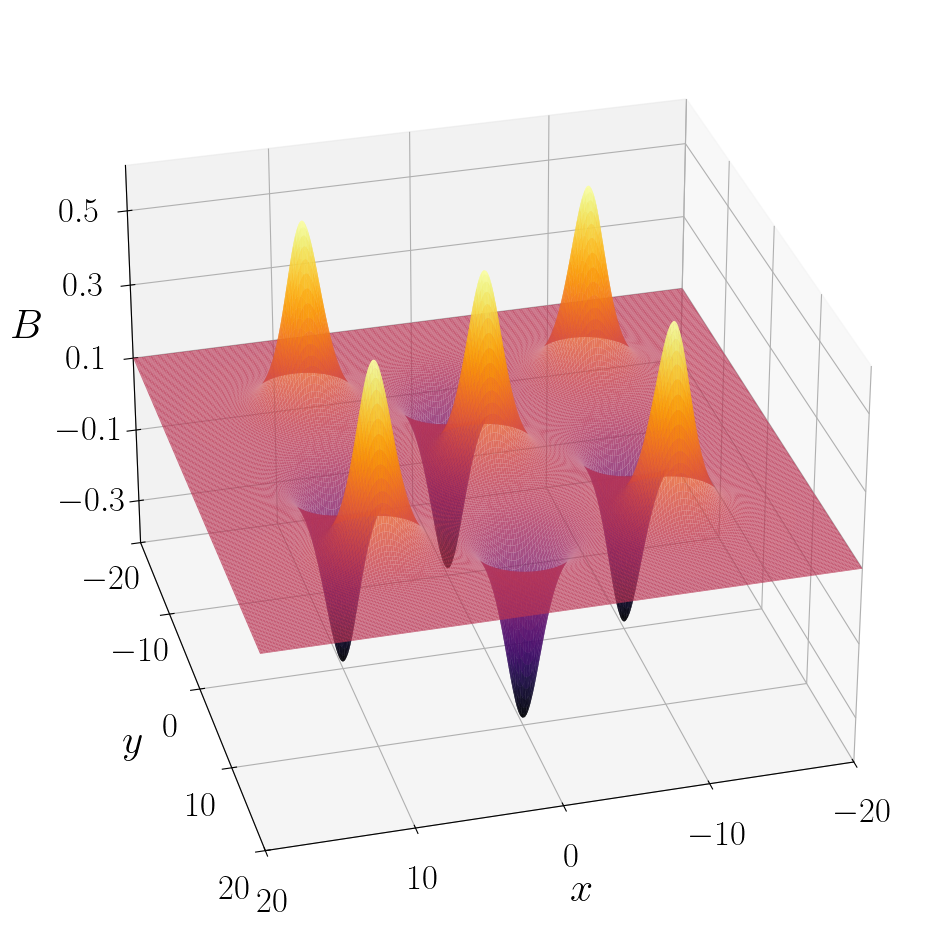}
        \caption{}
    \end{subfigure}
    
    \caption{Color-map and 3D plots of the magnetic field solution, $B^{(n_1, n_2)}$, for the iterations $(n_1,n_2) = (2,1)$ (a-b) and $(n_1,n_2) = (2,2)$ (c-d). Parameters are $(a_x^{(1)}, \beta_x, a_y^{(1)})=(5, 0, 5)$ (a-b) and  $(a_x^{(1)}, \beta_x, a_y^{(1)}, \beta_y)=(5, 0, 5, 0)$ (c-d). }

    \label{fig:6}
\end{figure}

To illustrate the construction, we present the solution $B^{(2,1)}$ for $(a_x^{(1)},\beta_x,a_y^{(1)})=(5,0,5)$ (figure~\ref{fig:6}(a-b)), and $B^{(2,2)}$ for $(a_x^{(1)},\beta_x,a_y^{(1)},\beta_y)=(5,0,5,0)$ (figure~\ref{fig:6}(c-d)). As expected, when the parameters satisfy $\{a_x^{(n_1)}\}=\{a_y^{(n_2)}\}$, the resulting lattice forms a rectangular array with $n_1+1$ columns and $n_2+1$ rows, in which vortices and antivortices alternate. More generally, the moduli space of a solution obtained after $n_1$ iterations of $\sigma_1$ and $n_2$ iterations of $\sigma_2$ is parametrized by $n_1+n_2$ moduli.

Interestingly, the iterative scheme highlights that the BPS-impurity equations (\ref{eq:2.7}) and (\ref{eq:2.8}), when coupled to mirror impurities, admit solutions belonging to different topological sectors. For example, the solutions with $(n_1,n_2)=(1,0)$ (v-av) and $(n_1,n_2)=(2,0)$ (v-av-v) have total topological charges $N=0$ and $N=1$, respectively. This behavior is expected, since the chosen impurities has nontrivial asymptotic profiles. In other words, they are not localized and therefore modify the boundary conditions of the fields. We return to the topological properties of these solutions in section~\ref{sec:5}.

\section{Radial impurities} 
\label{sec:4}

In this section, we present an alternative analytical approach to decouple eqs.~(\ref{eq:2.7})-(\ref{eq:2.8}). We search for radially symmetric solutions by choosing the impurity functions to depend only on the radial coordinate, i.e., $\sigma_1(x,y)=\sigma_1(r)$ and $\sigma_2(x,y)=\sigma_2(r)$. We then substitute the standard radial vortex ansatz, given by eqs.~(\ref{eq:3.9})-(\ref{eq:3.11}) with $(r,\theta)$ replacing $(\rho,\vartheta)$, into the BPS-impurity equations. As a result, they reduce to
\begin{align}
&\sigma_1(r)f' = \sigma_2(r) \frac{f}{r}(1-g), \label{eq:4.1} \\
&\sigma_1(r)\sigma_2(r)g' = \frac{r}{2}(1-f^2).\label{eq:4.2}
\end{align}

Having established the general framework, we now consider a single radial bump impurity, defined as an impurity function satisfying $\sigma_i\to1$ as $r\to\infty$, with an adjustable amplitude near the origin. The effect of such an impurity can be interpreted as either stretching or compressing the vortex, as we show below. To model these deformations, we choose $\sigma_1=1$ and
\begin{align}
&\sigma_2(r) \equiv \sigma(r) = 1 + \frac{\alpha r^{\beta}}{\cosh^2(r)}, \label{eq:4.3}
\end{align} 
where $\alpha$ and $\beta$ are free parameters, with $\beta\geq0$ to make sure that the impurity remains regular at the origin.

The particular case $\alpha=-2c$ and $\beta=0$ is a two-dimensional radial version of the bump impurity considered in ref.~\cite{manton2019}. In the corresponding one-dimensional model, the bump initially induces compression, followed by a slight stretching. As its strength increases further, the deformation quickly evolves into an antikink-kink pair.

\begin{figure}[t]
    \centering
    % Primeira linha (2 figuras)
    \begin{subfigure}{0.48\textwidth}
        \includegraphics[width=\linewidth]{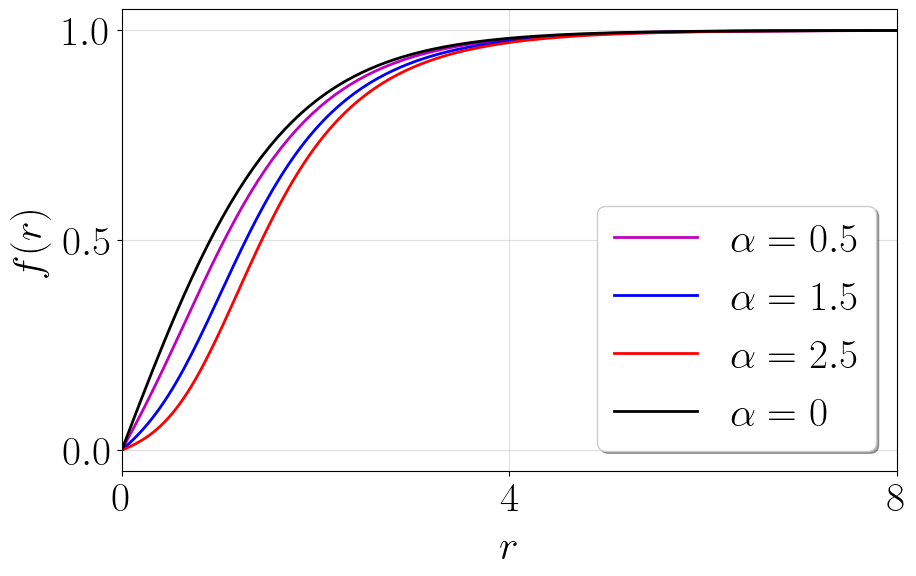}
        \caption{$\beta=0$}
    \end{subfigure}
    \hfill
    \begin{subfigure}{0.48\textwidth}
        \includegraphics[width=\linewidth]{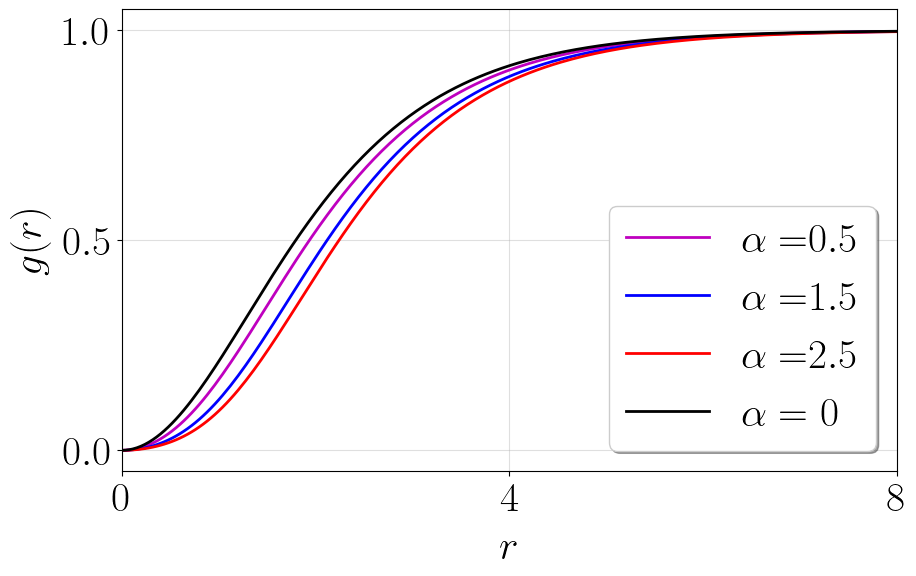}
        \caption{$\beta=0$}
    \end{subfigure}

    \vspace{0.01cm}

    \begin{subfigure}{0.48\textwidth}
        \includegraphics[width=\linewidth]{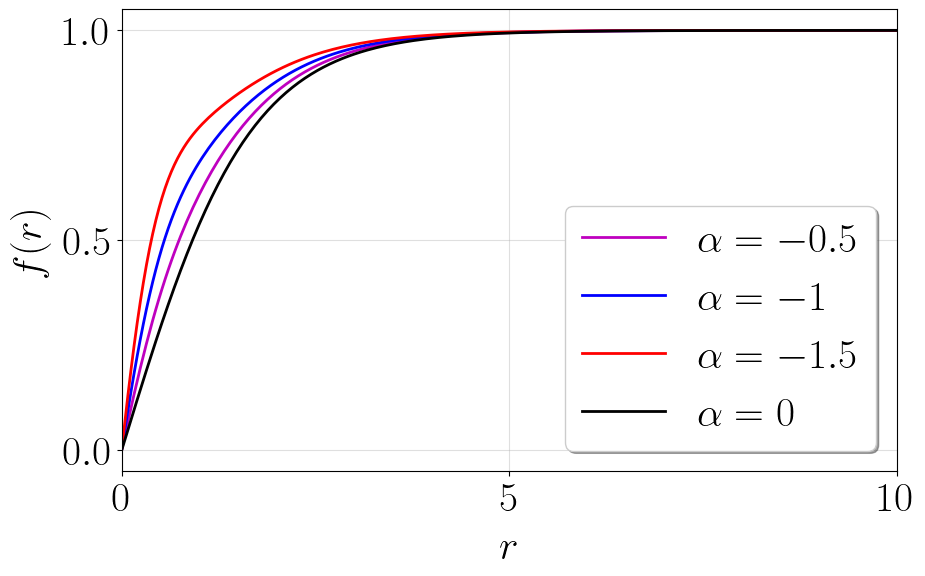}
        \caption{$\beta=1.5$}
    \end{subfigure}
    \hfill
    \begin{subfigure}{0.48\textwidth}
        \includegraphics[width=\linewidth]{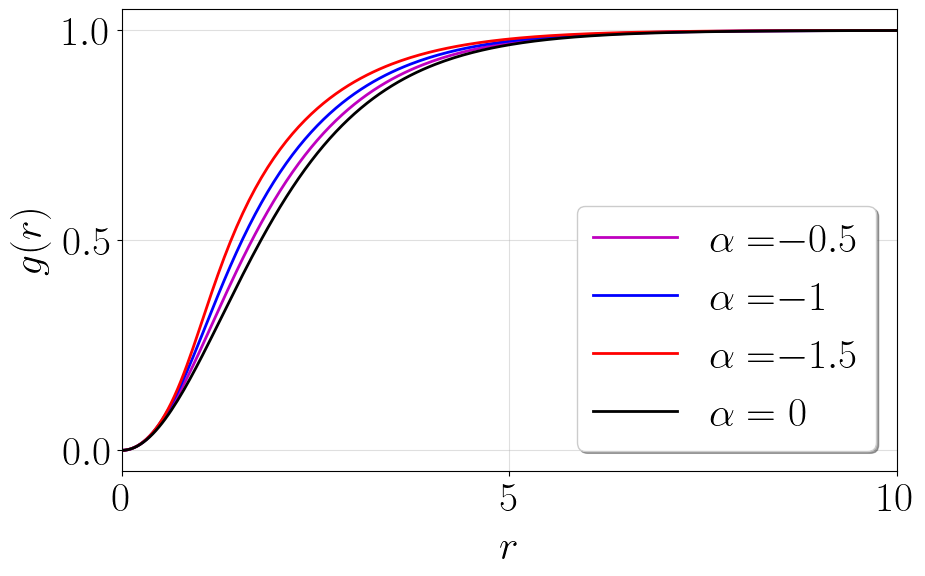}
        \caption{$\beta=1.5$}
    \end{subfigure}

    \caption{Numerical solutions for the functions $\{f(r),g(r)\}$ corresponding to stretched vortices (a-b), obtained for several nonnegative values of $\alpha$ with $\beta=0$, and compressed vortices (c-d), obtained for several nonpositive values of $\alpha$ with $\beta=1.5$.}

    \label{fig:7}
\end{figure}

Here, since v-av pairs cannot be obtained within a radially symmetric ansatz, we instead focus on spatially deformed vortex configurations. In particular, for $\alpha=0$, the solutions reduce to the standard Nielsen-Olesen vortex at critical coupling. For positive values of $\alpha$, however, the bump impurity stretches the vortex, as shown in figure~\ref{fig:7}(a-b). As $\alpha$ increases, the deformation becomes more pronounced and the functions $\{f(r),g(r)\}$ spread over a larger spatial region, particularly near the origin, gradually developing an ``S''-shaped profile. The parameter $\beta$ also increases the stretching, but mostly controls the smoothness of the profiles. In the examples shown in figure~\ref{fig:7}(a-b), we fix $\beta=0$ and consider several nonnegative values of $\alpha$, where we obtain behavior similar to that observed in the one-dimensional model.

The introduction of the power term into the impurity also gives rise to a wider class of compressed-vortex solutions. This point deserves further explanation. For $\beta=0$ and $\alpha\leq-1$, the impurity vanishes at a finite radius $r=r_0$, which, according to eqs.~(\ref{eq:4.1})-(\ref{eq:4.2}), imposes the conditions $f^2(r_0)=1$ and $f'|_{r=r_0}=0$. Consequently, compressed-vortex solutions exist only for the narrow interval $-1<\alpha<0$. By choosing $\beta>0$, this interval is extended, since the zeros of $\sigma(r)$ are shifted to more negative values of $\alpha$. As a result, a broader range of negative values of $\alpha$ gives rise to regular compressed-vortex solutions.

Figure~\ref{fig:7}(c-d) displays the solutions for $\beta=1.5$ and several nonpositive values of $\alpha$. As $\alpha$ decreases, the bump becomes more pronounced, leading to an increasingly compressed vortex. In this case, the first zero of $\sigma(r)$ appears only at the threshold value $\alpha_{\rm th}\approx-2.38$, beyond which the solutions no longer exhibit a vortex-like profile. As in the stretched-vortex case, both functions $f(r)$ and $g(r)$ remain monotonically increasing, since the bump impurity $\sigma(r)$ does not change sign.

An important consequence of the bump impurity is its effect on the internal structure of the resulting vortices. This can be observed through the magnetic field, which is given by
\begin{align}
&B(r)=\frac{g'}{r}=\frac{1}{2\sigma}(1-f^2). \label{eq:4.4}
\end{align}
For $\beta=0$, the impurity modifies the magnetic field at the origin in the form $B(0) = \frac{1}{2(1+\alpha)}$.
Equation~(\ref{eq:4.4}) further shows that a new zero of $B'(r)$ appears at a finite distance from the origin. 
In the regime $\alpha>0$ and $\beta=0$, this point corresponds to the global maximum of $B(r)$, producing a magnetic-field peak at a finite distance from the vortex center.

\begin{figure}[b!]
    \centering
    % Primeira linha (2 figuras)
    \begin{subfigure}{0.45\textwidth}
        \includegraphics[width=\linewidth]{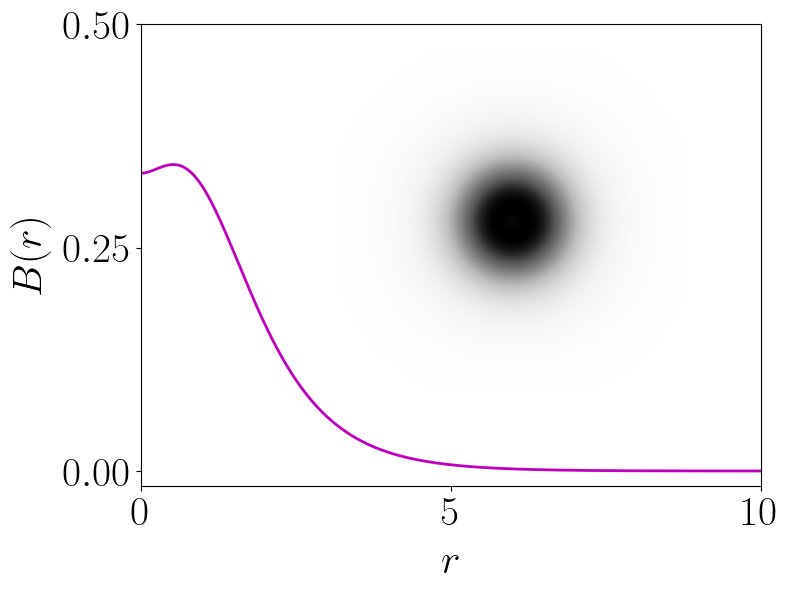}
        \caption{$(\alpha, \beta)=(0.5, 0)$}
    \end{subfigure}
    \hfill
    \begin{subfigure}{0.45\textwidth}
        \includegraphics[width=\linewidth]{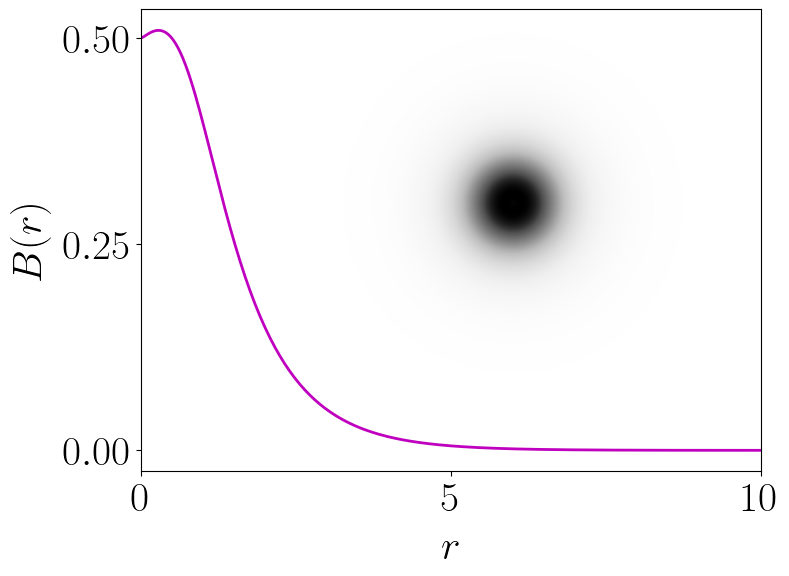}
        \caption{$(\alpha, \beta)=(-0.5, 1.5)$}
    \end{subfigure}

    \vspace{0.01cm}

    \begin{subfigure}{0.45\textwidth}
        \includegraphics[width=\linewidth]{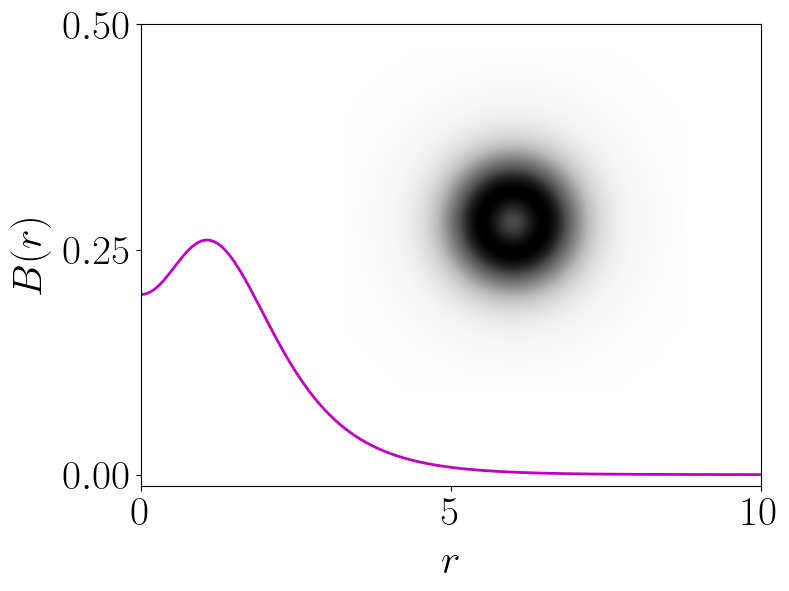}
        \caption{$(\alpha, \beta)=(1.5, 0)$}
    \end{subfigure}
    \hfill
    \begin{subfigure}{0.45\textwidth}
        \includegraphics[width=\linewidth]{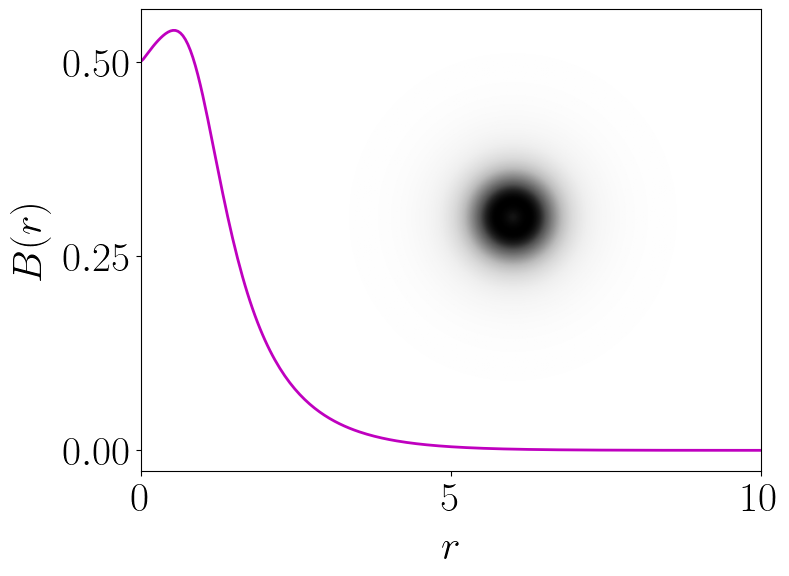}
        \caption{$(\alpha, \beta)=(-1, 1.5)$}
    \end{subfigure}

    \vspace{0.01cm}

    \begin{subfigure}{0.45\textwidth}
        \includegraphics[width=\linewidth]{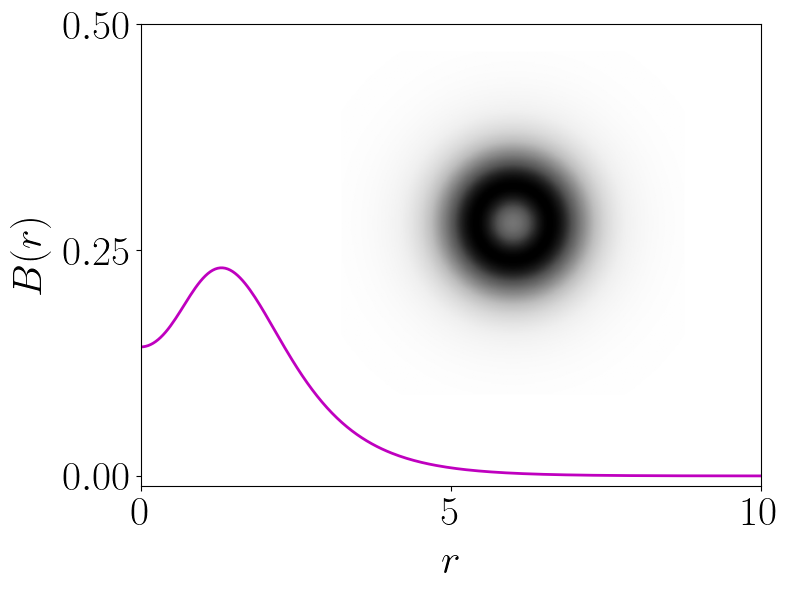}
        \caption{$(\alpha, \beta)=(2.5, 0)$}
    \end{subfigure}
    \hfill
    \begin{subfigure}{0.45\textwidth}
        \includegraphics[width=\linewidth]{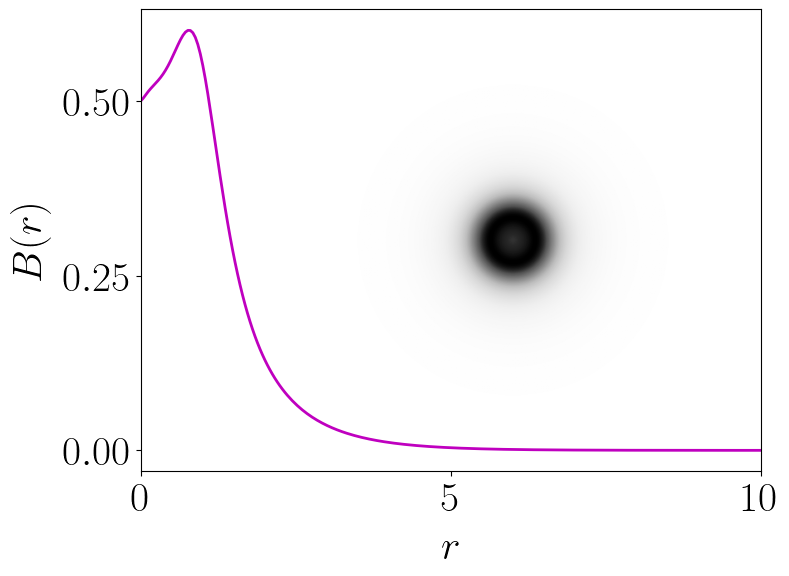}
        \caption{$(\alpha, \beta)=(-1.5, 1.5)$}
    \end{subfigure}
    
    \caption{Radial and planar $B(r)$ for stretched (a-c-e) and compressed (b-d-f) vortices. All planar sections have the same scale.}

    \label{fig:8}
\end{figure}

Figure~\ref{fig:8} shows the magnetic-field profiles for representative values of the parameters $(\alpha,\beta)$. The left panels (a, c, e) correspond to the stretched-vortex solutions presented in figure~\ref{fig:7}, while the right panels (b, d, f) show the compressed-vortex solutions. For $\beta>0$ and $\alpha<0$ (right panels), the value at the origin remains fixed, while the bump impurity shifts the position of the magnetic-field peak. In both stretched and compressed vortices, the central cavity becomes wider as 
$|\alpha|$ increases.

\begin{figure}[h!]
    \centering
    % Primeira linha (2 figuras)
    \begin{subfigure}{0.48\textwidth}
        \includegraphics[width=\linewidth]{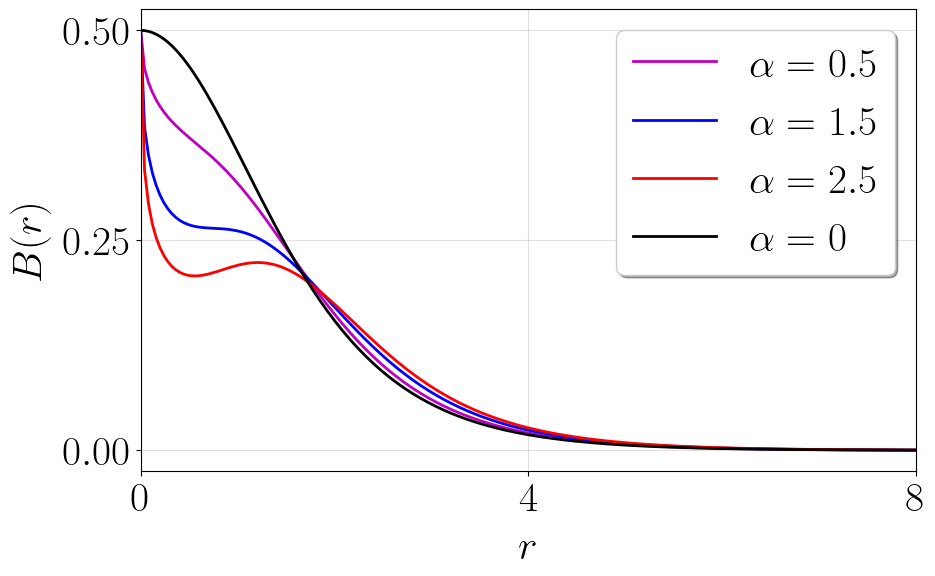}
        \caption{$\beta=0.5$}
    \end{subfigure}
    \hfill
    \begin{subfigure}{0.48\textwidth}
        \includegraphics[width=\linewidth]{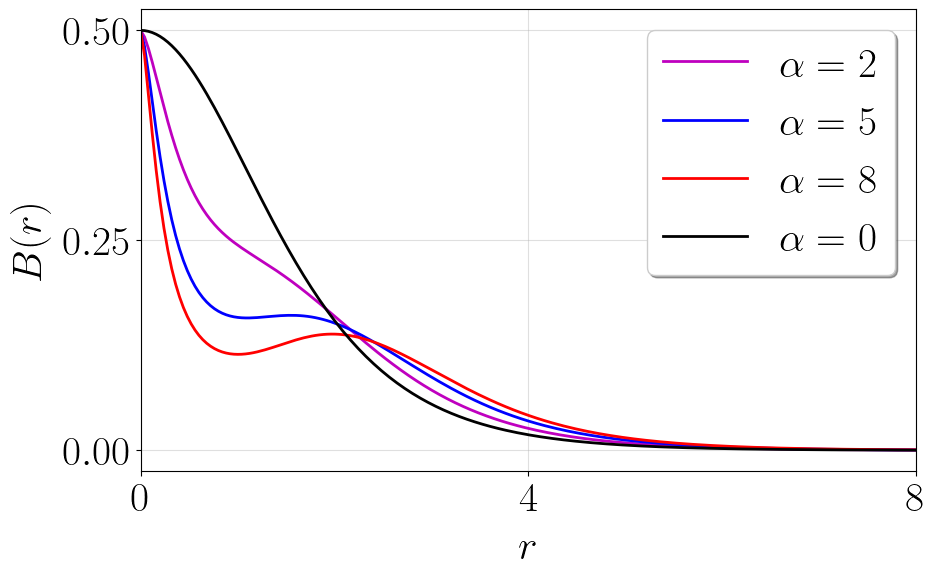}
        \caption{$\beta=1.5$}
    \end{subfigure}

    \caption{Magnetic field $B(r)$ profiles as a function of radius $r$ for several values of $\alpha$ and $\beta$.}

    \label{fig:9}
\end{figure}

Other magnetic-field profiles can be obtained by combining positive values of $\alpha$ and $\beta$. For $0<\beta<1$, the derivative $\sigma'(r)$ diverges to $-\infty$ at the origin. As a result, the magnetic field decreases sharply near the vortex center. As $\alpha$ increases, two additional extrema emerge, corresponding to a local minimum and a local maximum, as shown in figure~\ref{fig:9}(a) for $\beta=0.5$. Finally, for $\beta>1$, the impurity function and all of its derivatives remain regular, leading to a smoother magnetic-field profile near the origin. As in the previous case, increasing $\alpha$ results in profiles with three extrema, as shown in figure~\ref{fig:9}(b).

\begin{figure}[b!]
    \centering
    % Primeira linha (2 figuras)
    \begin{subfigure}{0.48\textwidth}
        \includegraphics[width=\linewidth]{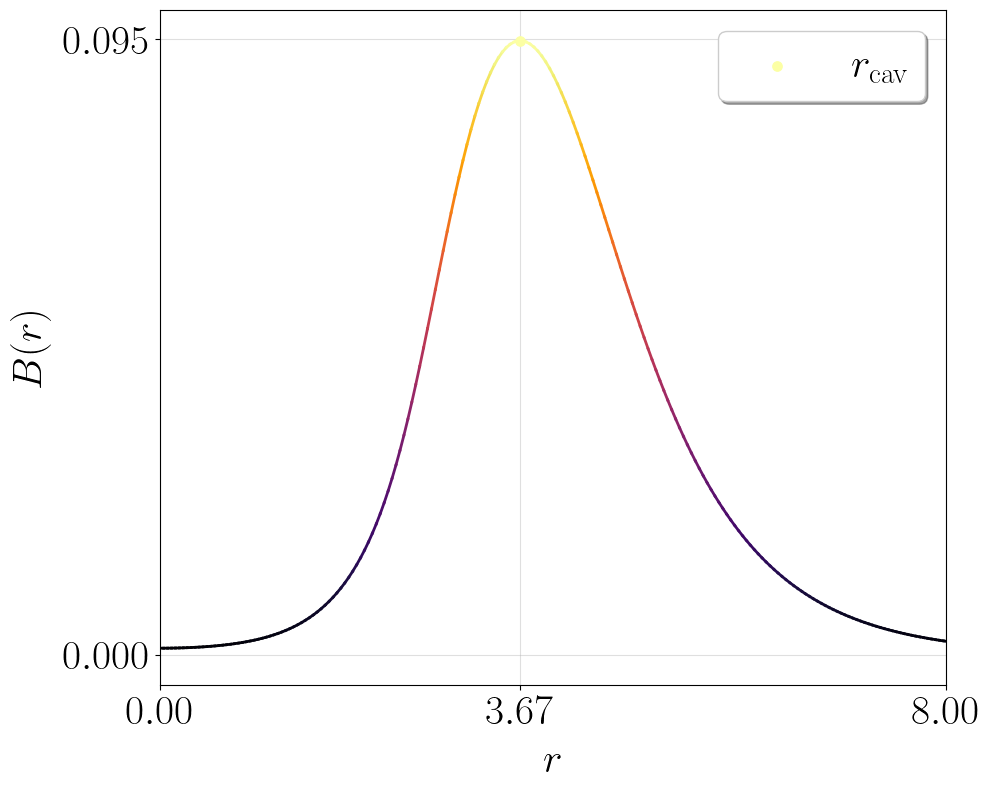}
        \caption{}
    \end{subfigure}
    \hfill
    \begin{subfigure}{0.48\textwidth}
        \includegraphics[width=\linewidth]{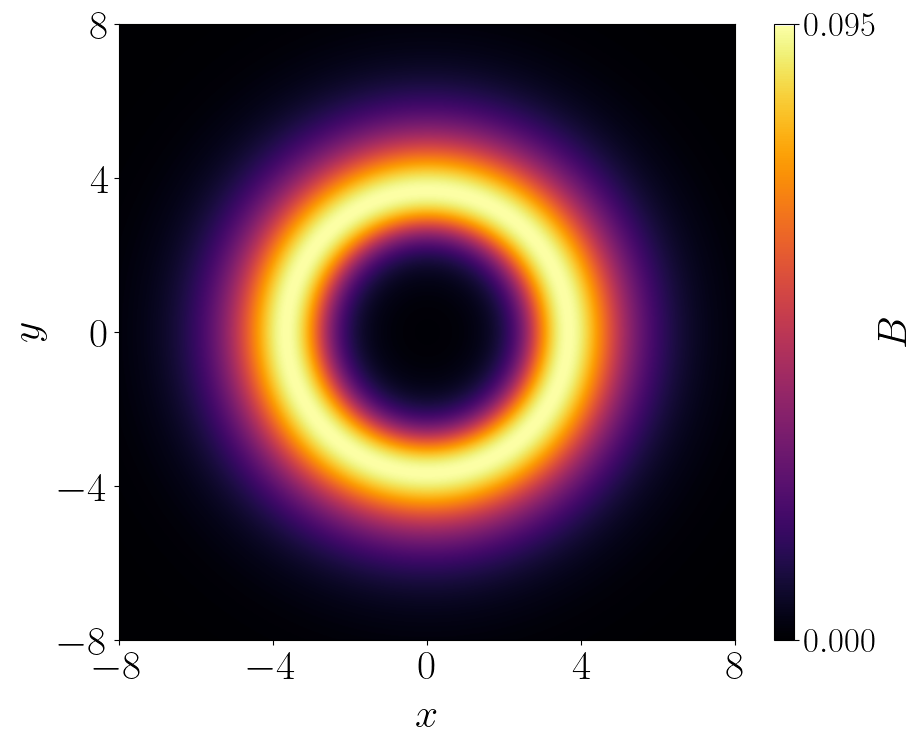}
        \caption{}
    \end{subfigure}

    \caption{1D representation (a) and color-map (b) of the magnetic field of the hollow vortex in the presence of the radial bump-impurity. We fix $\alpha=500$ and $\beta=0$.}

    \label{fig:10}
\end{figure}

A final interesting case is the limit $\alpha\to\infty$ with $\beta=0$. As discussed previously, for $\alpha>0$ and $\beta=0$ the magnetic field at the origin is given by $B(0)=1/[2(1+\alpha)]$, which vanishes in the limit $\alpha\to\infty$. Consequently, the magnetic field develops a cavity around the vortex center, where the magnetic energy density becomes strongly suppressed. Figure~\ref{fig:10} shows the corresponding magnetic-field profile for $\alpha=500$. The cavity has an estimated radius of $r_{\rm cav}\approx3.67$, while the magnetic field reaches a maximum value of $B_{\rm max}\approx0.93$, as indicated in figure~\ref{fig:10}. From the perspective of the magnetic-field distribution, this configuration can be considered as a hollow vortex, showing that how radial bump impurities can generate nontrivial internal structures in $(2+1)$-dimensional vortex solutions.

\section{Lagrangian formalism} \label{sec:5}

In this section, we construct a Lagrangian formalism whose static Bogomol'nyi equations are given by eqs.~(\ref{eq:2.7})-(\ref{eq:2.8}). As a starting point, we consider the energy functional
\begin{align}
    E &= \frac{1}{2}\int d^2x \left\{ \left|\sigma_1 D_1 \phi + i\sigma_2 D_2 \phi \right|^2  + \left[\sigma_1 \sigma_2 B + \frac{1}{2}\left(|\phi|^2 - 1\right)\right]^2 \right\} + \frac{1}{2}\int d^2x \; B \; \nonumber \\ 
    & \geq  \frac{1}{2}\int d^2x \; B,   \label{eq:5.1}
\end{align} 
which follows the standard Bogomol'nyi construction, where the energy is written as a sum of squared first-order equations plus a boundary term.

Notice that, the second integral has the same form as the standard topological term of the Abelian-Higgs model. However, the presence of the impurity, particularly those with nontrivial asymptotic behavior considered in this work, modifies the asymptotic boundary conditions of the fields and, consequently, their topological sector. As a result, the value of the integral generally depends on the choice of the impurity functions $\sigma_i$ $(i=1,2)$. For this reason, we leave the topological term in its integral form.

Proceeding further, we rewrite the energy as
\begin{align}
    E = &\frac{1}{2}\int d^2x \left\{
    \sigma_1^2 \sigma_2^2 B^2 + \sigma_i^2|D_i \phi|^2
    + \frac{1}{4}\left(|\phi|^2 - 1\right)^2 
    + B (1 - \sigma_1 \sigma_2) \nonumber \right. \\
    &\left. + i \sigma_1 \sigma_2 \left[ \partial_1\left(\bar{\phi} D_2 \phi \right) - \partial_2\left(\bar{\phi} D_1 \phi\right) 
     \right]
    \right\} , 
    \label{eq:5.2}
\end{align} 
which can be considered as the potential energy $V$ of the impurity model.
Hence, the corresponding static Lagrangian is given by 
\begin{align}
    L_{\text{static}} = -V = &-\frac{1}{2}\int d^2x \left\{
    \sigma_1^2 \sigma_2^2 B^2 + \sigma_i^2|D_i \phi|^2 
    + \frac{1}{4}\left(|\phi|^2 - 1\right)^2 
    + B (1 - \sigma_1 \sigma_2) \nonumber \right. \\
    &\left. + i \sigma_1 \sigma_2 \left[ \partial_1\left(\bar{\phi} D_2 \phi \right) - \partial_2\left(\bar{\phi} D_1 \phi\right) 
     \right]
    \right\} . \label{eq:5.3}
\end{align}

Moreover, we define the kinetic energy of the model as the standard one, in order to maintain Gauss's law unaffected by the presence of the impurity.
Thus, it takes the following form
\begin{align}
     T = \int d^2 x\; \left( \frac{1}{2}E_iE_i + \frac{1}{2}|D_0 \phi|^2 \right) ,  \label{eq:5.4}
\end{align} 
where $E_i=\partial_0 A_i-\partial_i A_0$ are the components of the electric field. The complete Lagrangian is then given by
\begin{align}
     L = T + L_{\text{static}} , \label{eq:5.5}
\end{align} 
where $T$ and $L_{\text{static}}$ are defined in eqs.~(\ref{eq:5.4}) and (\ref{eq:5.3}), respectively.

Recalling the conditions in eq.~(\ref{eq:3.19}), we require the terms multiplying $\sigma_i$ $(i=1,2)$ to vanish at the points where the impurity functions diverge. With the Lagrangian (\ref{eq:5.5}), it also becomes possible to investigate the second-order dynamics of the fields in the presence of such impurities.

\section{The geodesic approximation} \label{sec:6}

As discussed in section \ref{sec:3}, for a mirror impurity such as
$\sigma_1(x)=\coth x$, the v-av pair belongs to a one-dimensional moduli space\footnote{Once again, we neglect the
translational moduli.}, which we parametrize by $\alpha_x$. To describe
the dynamics within the moduli-space approximation, we promote this
modulus to a time-dependent collective coordinate,
$\alpha_x=\alpha(t)$ where we omit the subscript $x$ for simplicity. 
For sufficiently small velocities, the fields are assumed to remain
close to the family of BPS configurations, with their time dependence
entering only through $\alpha(t)$. Since all configurations along this family have the same static energy, we can model the resulting small-velocity dynamics as geodesic motion on the moduli space.

The time dependence of the fields is introduced through the modulus
$\alpha(t)$. Thus, recalling the fields given in eqs.~(\ref{eq:3.14})-(\ref{eq:3.15}), we write
\begin{align}
&\phi(u(x;\alpha), y, t) =\phi(u(x;\alpha(t)), y), \label{eq:6.6} \\
&A_{i}(u(x;\alpha), y, t) = A_{i}(u(x;\alpha(t)), y) \label{eq:6.7} , \;\;\;i = 1, 2 . 
\end{align} 
Considering the Lagrangian (\ref{eq:5.5}) we developed in section \ref{sec:5}, for this BPS dynamics, the potential part remains constant throughout the motion. 
Up to an irrelevant additive constant, the reduced Lagrangian is therefore
purely kinetic and takes the form
\begin{align}
    \mathcal{L}_{\text{red}} = \frac{1}{2}M(\alpha)\dot{\alpha}^2, \label{eq:6.8}
\end{align} 
where $M(\alpha)$ is the metric function of the one-dimensional moduli
space. To determine $M(\alpha)$, we substitute the time-dependent fields $\{\phi, A_{\mu}\}$
in eqs.~(\ref{eq:6.6})-(\ref{eq:6.7}) into the kinetic energy $T$ given
by eq.~(\ref{eq:5.4}). After evaluating the
corresponding contributions, we obtain
\begin{align} \hspace{-0.18cm}
M(\alpha) = \int d^2x \left(\frac{\partial u}{\partial \alpha} \right)^2 \left\{
\left( \frac{f' u}{\rho} \right)^2 +
\left( \frac{f y}{\rho^2} \right)^2 +
\left[ \frac{y u (g'\rho - 2g)}{\sigma_1 \rho^4} \right]^2 + 
\left[ \frac{u^2 (g'\rho - 2g)}{\rho^4} + \frac{g}{\rho^2} \right]^2
\right\}. \label{eq:6.9}
\end{align}

It is straightforward to obtain the geodesic equation that follows from eqs. (\ref{eq:6.8})-(\ref{eq:6.9}). Its solutions satisfy the following differential equation, 
\begin{align}
    \dot{\alpha}(t) \equiv v(t) = v_0 \sqrt{\frac{M(\alpha_0)}{M(\alpha(t))}}, \label{eq:6.10}
\end{align} 
where $(\alpha_0, v_0)$ are the initial conditions for the position and velocity of the pair.

\begin{figure}[t!]
    \centering
    % Primeira linha (2 figuras)
    \begin{subfigure}{0.49\textwidth}
        \includegraphics[width=\linewidth]{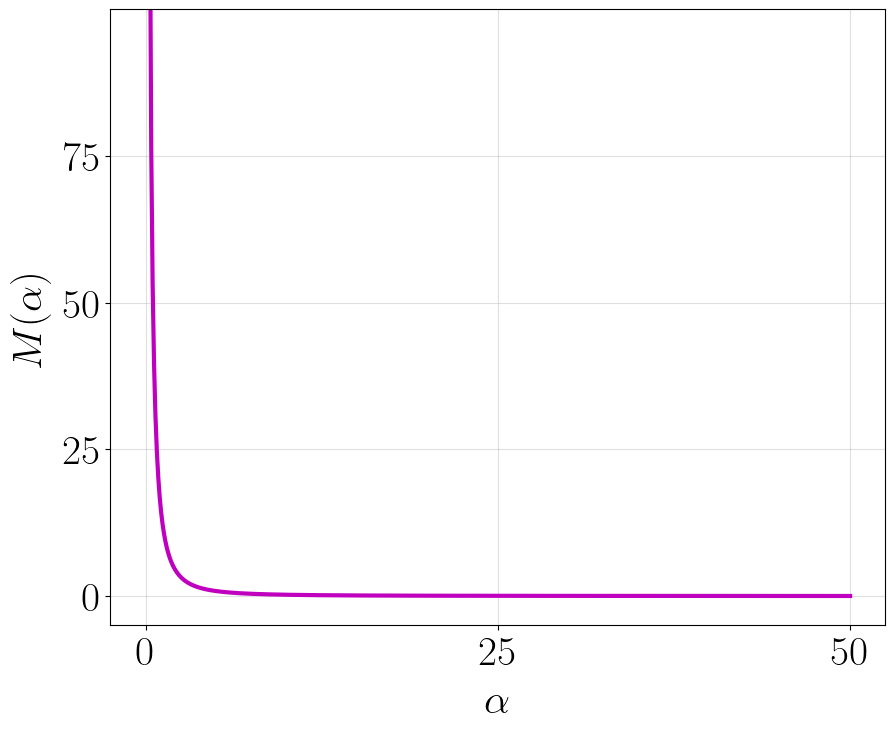}
        \caption{}
    \end{subfigure}
    \hfill
    \begin{subfigure}{0.48\textwidth}
        \includegraphics[width=\linewidth]{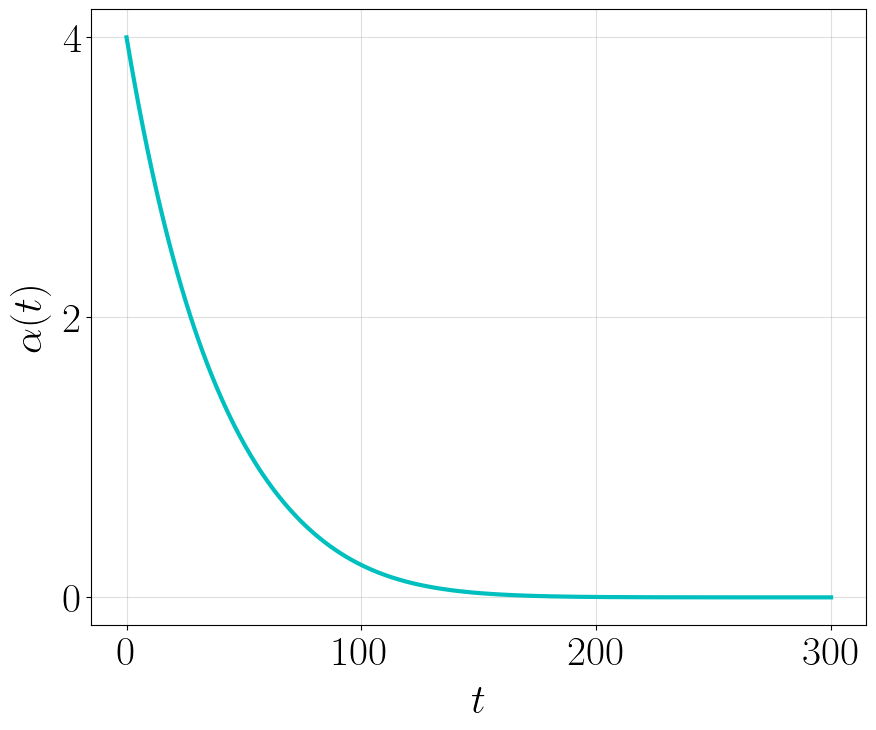}
        \caption{}
    \end{subfigure}
    
    \caption{(a) Moduli-space metric $M(\alpha)$ for a single
    mirror impurity, $\sigma_1(x)=\coth x$, as a function of the
    collective coordinate $\alpha$. (b) Corresponding solution
    $\alpha(t)$ of the geodesic equation for the initial conditions
    $(\alpha_0,v_0)=(4,-0.1)$.}

    \label{fig:6.2}
\end{figure}

For the mirror impurity $\sigma_1(x)=\coth x$, figure~\ref{fig:6.2}(a) shows the dependence of the moduli-space metric
$M(\alpha)$ on the collective coordinate $\alpha$. Predictably, as the vortex moves away from the impurity, the metric becomes constant, vanishing for $\alpha \to \infty$, similarly to the one-dimensional impurity model \cite{adam2019}. In the limit $\alpha \to 0$, the metric diverges due to the singular impurity.

Starting from a well-separated v-av pair with a small initial velocity, the system evolves toward smaller values of $\alpha$
along a geodesic of the moduli space due to the force generated by the curvature of the moduli space. As $\alpha\to0$, the metric $M(\alpha)$ diverges and, according to eq.~(\ref{eq:6.10}), the velocity $\dot{\alpha}$ tends to zero.
Consequently, the system approaches $\alpha=0$ only asymptotically,
taking an infinite amount of time to reach the configuration corresponding to complete v-av annihilation, as discussed in section~\ref{sec:3.A}.  This behavior is shown in figure~\ref{fig:6.2}(b), which shows the solution of the moduli space equations $\alpha(t)$ for the
initial conditions $\alpha_0=4$ ($d\approx4$ between the pair) and $v_0=-0.1$.

The geodesic description of the collision is subject to limitations arising from the singular nature of the mirror impurity,
$\sigma_1(x)=\coth x$. In particular, the singularity at $x=0$ effectively introduces a ``wall'' at the location of the impurity, constraining the magnetic field to vanish along this line.

As discussed previously, in the Abelian-Higgs model without impurities,
the first-order BPS equations do not admit configurations with both vortices and antivortices. Consequently, v-av collisions in the standard Abelian-Higgs theory must be studied using the full second-order field equations. The present impurity model provides an alternative framework in which a v-av pair can be described within a family of BPS configurations, allowing its approach toward annihilation to be studied through moduli-space dynamics. This description, however, does not capture the complete annihilation process of the standard theory. There, the energy initially carried by the BPS v-av pair is redistributed through
the emission of radiation. Such effects lie beyond the geodesic approximation, which describes motion only within the family of BPS configurations. Therefore, although the present construction provides
a BPS description of the approach toward v-av annihilation, a complete dynamical description of the Abelian-Higgs model using the present framework requires going beyond the moduli-space approximation.

\subsection{Beyond the geodesic approximation}

The geodesic approximation developed above describes the v-av collision as motion along a family of BPS configurations. In particular, the analytical continuation of the modulus $\alpha_x$ introduced in section~\ref{sec:3.A} provides a continuation of this family toward the annihilation configuration. A natural question is whether this moduli-space trajectory reproduces the full-field dynamics. Addressing this question would require solving the second-order field equations and comparing the resulting evolution with the geodesic trajectory.

The coordinate transformation introduced in section~\ref{sec:3.A} maps the BPS-impurity equations onto the standard BPS equations of the Abelian-Higgs model. However, this equivalence holds only at the level of the first-order BPS equations. The full Lagrangian (\ref{eq:5.5}) does not reduce to the impurity-free Abelian-Higgs Lagrangian under the same transformation, since the impurity dependence remains in the action. Therefore, the full dynamics must be studied directly from the second-order equations of motion of the impurity model.

However, there appears some complications in solving the second-order full dynamics due to the singular mirror impurity. For a general impurity $\sigma_1=\sigma_1(x)$ with $\sigma_2=1$, the Euler-Lagrange equations derived from the Lagrangian~(\ref{eq:5.5}) in the temporal gauge $A_0=0$, are
\begin{align}
&D_0 D_0 \phi-\sigma_1^2 D_1 D_1 \phi-D_2 D_2 \phi-\sigma_1^{\prime}\left(2 \sigma_1 D_1 \phi+i D_2 \phi\right)+\frac{1}{2}\left(|\phi|^2 - 1 \right) \phi = 0,  
\label{eq:6.11} \\
&\partial_0^2 A_1+\sigma_1^2\partial_2 B+\sigma_1^2 J_1=0, 
\label{eq:6.12} \\
&\partial_0^2 A_2-\sigma_1^2\partial_1 B-2 \sigma_1\sigma_1' B-\frac{\sigma_1'}{2}\left(|\phi|^2-1\right)+J_2=0, \label{eq:6.13}
\end{align}
 where, once again, $J_i = \frac{i}{2}(\overline{\phi}D_i \phi - \phi \overline{D_i \phi})$ is the usual electric current of the Abelian-Higgs model in the absence of impurities.

For the mirror impurity considered throughout this section, $\sigma_1(x)=\coth x$, both $\sigma_1$ and its derivative are singular at $x=0$. Regularity of the second-order equations therefore imposes additional conditions on the fields along the impurity line. To avoid divergences in eqs.~(\ref{eq:6.11})-(\ref{eq:6.13}) lead to the following conditions
on the line $x=0$:
%\begin{align}
%&D_1D_1\phi\big|_{x=0}=D_1\phi\big|_{x=0}= D_2\phi\big|_{x=0}=0,
%\label{eq:6.14}\\
%&|\phi(0,y)|^2=1, \qquad B\big|_{x=0}=
%\partial_1B\big|_{x=0}=\partial_2B\big|_{x=0}=0.
%\label{eq:6.15}
%\end{align}
\begin{align}
&D_1\phi\big|_{x=0}=0,
\qquad
D_1D_1\phi\big|_{x=0}
=
-iD_2\phi\big|_{x=0},
\label{eq:6.14}
\\
&B\big|_{x=0}=0,
\qquad
\partial_2B\big|_{x=0}=0,
\qquad
\partial_1B\big|_{x=0}
=
-\frac{1}{2}
\left(
|\phi(0,y)|^2-1
\right).
\label{eq:6.15}
\end{align}
These conditions supplement the usual boundary conditions, leaving the corresponding dynamical problem overdetermined and thus making direct time integration impossible. This issue could be avoided by choosing a non-divergent impurity. However, such an approach lies beyond the scope of the present manuscript, since non-divergent impurities do not admit v-av solutions.

\section{Conclusion}
\label{sec:conc}

In this paper, we have introduced a novel family of BPS vortex solutions
coupled to scalar impurities $\sigma_i(\mathbf{x})$ $(i=1,2)$. We showed
that, for a suitable class of impurity functions, the corresponding Bogomol'nyi equations can be mapped through a coordinate transformation onto the standard BPS equations of the impurity-free Abelian-Higgs model, eqs.~(\ref{eq:2.5})-(\ref{eq:2.6}). 
In particular, we showed that impurities of the $\coth$ type generate
vortex-antivortex pairs in which the impurity is located at the midpoint
between the two solitons. This reflection-like property motivates the term \emph{mirror impurities}.

We further showed that an appropriate choice of modulus allows the v-av pair to approach complete annihilation along the BPS
moduli space. Despite the limitations of the construction discussed above, reproduces the typical annihilation observed in numerical simulations of v-av collisions in the full second-order theory \cite{rebbi1992}.

We then extended the mirror-impurity construction through an iterative
scheme involving the two independent sequences $\sigma_1^{(n_1)}$ and $\sigma_2^{(n_2)}$. Iterations along a single spatial direction generate chains of alternating vortices and
antivortices. Up to the second iteration, the construction remains
analytical and gives a vortex-antivortex-vortex configuration
parametrized by two moduli. Varying these moduli describes distinct BPS
processes in which a v-av pair annihilates, leaving a single vortex. More generally, performing $n_1$ and $n_2$ iterations
along the two spatial directions generates two-dimensional arrays of
alternating vortices and antivortices, including chessboard-like v-av lattices.

In section~\ref{sec:4}, we extended our construction to radially
symmetric solutions by decoupling eqs.~(\ref{eq:2.7})--(\ref{eq:2.8})
using a radial ansatz. We introduced a two-dimensional radial version
of the bump impurity considered in previous works, with an additional
power-law term. Depending on the impurity parameters, we obtained both
stretched and compressed vortices.
We also showed that the radial impurity can change the internal magnetic structure of the vortex. By exploring different regions of the parameter space, we found magnetic-field profiles with additional extrema and central cavities. In particular, for a strong bump impurity, the magnetic field becomes strongly suppressed at the vortex center, leading to what we call a \emph{hollow vortex}.

Starting from the BPS-impurity equations, we also constructed a Lagrangian formulation for the model. This allowed us to derive the
moduli-space metric and study the low-velocity dynamics of the v-av pair in the presence of a mirror impurity. Within the geodesic approximation, the pair approaches $\alpha=0$, corresponding to the annihilation configuration, only asymptotically. We also discussed the limitations of this
description, which are closely related to the singular nature of the
mirror impurity. In particular, the singularity imposes additional
conditions on the fields, making a direct numerical study of the full
second-order dynamics impossible.

Several directions remain open for future work. It would be interesting
to study the spectral properties of the mirror-impurity solutions and
to investigate whether similar constructions can be developed for
Chern-Simons vortices. Another natural direction is to study the
dynamics of the vortices generated by the radial bump impurities beyond
the BPS approximation, in particular to understand how their modified
internal structure affects vortex interactions and scattering.

\section*{Acknowledgements}
AM acknowledges financial support from CNPq (Conselho Nacional de
Desenvolvimento Científico e Tecnológico), Grant No.~306295/2023-7,
and from CAPES (Coordenação de Aperfeiçoamento de Pessoal de Nível
Superior). JPSN acknowledges financial support from FACEPE (Fundação
de Amparo à Ciência e Tecnologia do Estado de Pernambuco) through
scholarship BPG-1014-1.05/25.

\end{document}